\documentstyle[12pt,axodraw,epsf]{article}
\catcode`@=11
\newcount\@tempcntc
\def\@citex[#1]#2{\if@filesw\immediate\write\@auxout{\string\citation{#2}}\fi
  \@tempcnta\z@\@tempcntb\m@ne\def\@citea{}\@cite{\@for\@citeb:=#2\do
    {\@ifundefined
       {b@\@citeb}{\@citeo\@tempcntb\m@ne\@citea\def\@citea{,}{\bf ?}\@warning
       {Citation `\@citeb' on page \thepage \space undefined}}%
    {\setbox\z@\hbox{\global\@tempcntc0\csname b@\@citeb\endcsname\relax}%
     \ifnum\@tempcntc=\z@ \@citeo\@tempcntb\m@ne
       \@citea\def\@citea{,}\hbox{\csname b@\@citeb\endcsname}%
     \else
      \advance\@tempcntb\@ne
      \ifnum\@tempcntb=\@tempcntc
      \else\advance\@tempcntb\m@ne\@citeo
      \@tempcnta\@tempcntc\@tempcntb\@tempcntc\fi\fi}}\@citeo}{#1}}
\def\@citeo{\ifnum\@tempcnta>\@tempcntb\else\@citea\def\@citea{,}%
  \ifnum\@tempcnta=\@tempcntb\the\@tempcnta\else
   {\advance\@tempcnta\@ne\ifnum\@tempcnta=\@tempcntb \else \def\@citea{--}\fi
    \advance\@tempcnta\m@ne\the\@tempcnta\@citea\the\@tempcntb}\fi\fi}
\catcode`@=12

\def\theequation{\arabic{section}.\arabic{equation}}

\begin{document}

\begin{flushright}
MPI/PhT/97-30\\
hep-ph/9707235\\
May 1997
\end{flushright}

\begin{center}
{\LARGE {\bf CP Violation and Baryogenesis due to}}\\[0.4cm]
{\LARGE {\bf Heavy Majorana Neutrinos}}\\[2.4cm]
{\large Apostolos Pilaftsis}\footnote[1]{e-mail address: 
pilaftsi@mppmu.mpg.de}\\[0.4cm]
{\em Max-Planck-Institut f\"ur Physik, F\"ohringer Ring 6, 80805 
Munich, Germany}
\end{center}
\vskip1.4cm 
\centerline{\bf ABSTRACT} 
We analyze the scenario of baryogenesis through leptogenesis induced by 
the out-of-equilibrium decays of heavy Majorana neutrinos and pay
special attention to CP violation.  Extending a recently proposed
resummation formalism for two-fermion mixing to decay amplitudes, we
calculate the resonant phenomenon of CP violation due to the mixing of
two nearly degenerate heavy Majorana neutrinos.  Solving numerically
the relevant Boltzmann equations, we find that the isosinglet Majorana
mass may range from 1 TeV up to the grand unification scale, depending
on the mechanism of CP violation and/or the flavour structure of the
neutrino mass matrix assumed.  Finite temperature effects and possible
constraints from the electric dipole moment of electron and other
low-energy experiments are briefly discussed.\\[0.3cm] 
PACS no: 98.80.Cq

\newpage

\setcounter{equation}{0}
\section{Introduction}

Based on the assumption that the Universe was created initially in a
symmetric state with a vanishing baryon number $B$, Sakharov
\cite{ADS} derived the three known necessary conditions that may
explain the small baryon-to-photon ratio of number densities
$n_B/n_\gamma = (4-7)\times 10^{-10}$, which is found by present
observations. The first necessary ingredient is the existence of
$B$-violating interactions. With the advent of grand unified theories
(GUT's), this requirement can naturally be fulfilled at very
high-energy scales \cite{MY}, through the decay of super-heavy bosons
with masses near to the grand unification scale $M_X\approx 10^{15}$
GeV.  However, such a solution to the baryon asymmetry in the Universe
(BAU) faces some difficulties. In fact, Sakharov's second requirement
for generating the BAU prescribes that, by the same token, the
$B$-violating interactions should violate the discrete symmetries of
charge conjugation (C) and that resulting from the combined action of
charge and parity transformations (CP). One major drawback of the
solution suggested is that minimal scenarios of grand unification
generally predict very small CP violation, since it occurs at very
high orders in perturbation theory. Therefore, one has to rely on
no-minimal representations of GUT's in order to obtain appreciable CP
violation.  Furthermore, experiments on the stability of the proton
put tight constraints on the masses of the GUT bosons mediating $B$
violation and their couplings to the matter.

The most severe limits on scenarios for baryogenesis, however, come
from Sakharov's last requirement that the $B$- and CP-violating
interactions must be out of thermal equilibrium \cite{KW,EWK&SW}. In
the Standard Model (SM), the sum of $B$ and the lepton number $L$,
$B+L$, is violated anomalously \cite{sphal} through topologically
extended solutions, known as sphalerons.  In contrast to $B+L$
non-conservation, sphalerons preserve the quantum number $B-L$. The
authors in \cite{KRS} have found that $B+L$ anomalous violation may be
large at high temperatures $T$ above the critical temperature $T_c$ of
the electroweak phase transition. For values of $T$ not much larger
than the $W$-boson mass, $M_W$, {\em i.e.}\ $T>200$ GeV, up to
temperatures of $T\simeq 10^{12}$ GeV, the anomalous $B+L$ rate may
exceed the expansion rate of the Universe \cite{AMcL,BS}.  Therefore,
any primordial BAU generated at the GUT scale should not rely on
$B+L$-violating operators, since sphalerons being in thermal
equilibrium will then wash it out. The latter appears to be a generic
feature of most GUT's, where $B-L$-violating terms may be suppressed
against $B+L$-violating interactions, thus leading to the net effect
of a vanishing BAU.  On the other hand, it was suggested \cite{KRS}
that the same anomalous $B+L$ electroweak interactions may also be
utilized to produce the observed excess in $B$ during a first-order
electroweak phase transition. Given the fact that the experimental
lower mass bound of the Higgs boson $H$ is about $M_H>80$ GeV, this
scenario of electroweak baryogenesis must now be considered to be
rather improbable to explain the observed BAU \cite{ADD,RS} within the
minimal SM.

It is therefore important to note that baryogenesis not only provides
the strongest indication against the completeness of the SM but also
poses limits on its possible new-physics extensions. An attractive
scenario that may lead to a consistent solution to the problem of the
BAU is the one proposed by Fukugita and Yanagida \cite{FY},\footnote[1]{ 
The authors in \cite{KRS} presented an analogous scenario, which was, 
however, based on anomalous electroweak decays of exotic Dirac leptons
into quarks.} in which the baryon number is generated by 
out-of-equilibrium $L$-violating decays of heavy Majorana neutrinos
$N_i$ with masses $m_{Ni} \gg T_c$.  Moreover, it was argued \cite{FY}
that the excess in $L$ will then be converted into the desired excess
in $B$ by means of $B+L$-violating sphaleron interactions, which are
in thermal equilibrium above the critical temperature $T_c$.  Many
studies have been devoted to this mechanism of baryogenesis through
leptogenesis \cite{MAL,CEV,epsilonprime,LS1,Paschos} over the last
years.

Evidently, possible mechanisms for enhancing CP violation play
a decisive r\^ole in understanding the BAU. In \cite{FY}, the
necessary CP violation in heavy Majorana neutrino decays results from
the interference between the tree-level graph and the absorptive part
of the one-loop vertex.  Since CP violation originates entirely from
the decay amplitude in this case, we shall attach the characterization
of the $\varepsilon'$ type to this kind of CP violation, thereby
making contact with the terminology known from the $K^0\bar{K}^0$
system \cite{reviewCP}. The $\varepsilon'$-type CP violation was
discussed extensively in the literature \cite{MAL,CEV,epsilonprime}.
Provided all Yukawa couplings of the Higgs fields to $N_i$ and the
ordinary lepton isodoublets are of comparable order
\cite{MAL,epsilonprime}, baryogenesis through the $\varepsilon'$-type
mechanism requires very heavy Majorana neutrinos with masses not much
smaller than $10^7-10^8$ GeV.  If an hierarchical pattern for the
Yukawa couplings and the heavy Majorana neutrino masses is assumed
\cite{MAL,CEV}, the above high mass bound may be lifted and the
lightest heavy neutrino can have a mass as low as 1 TeV. Taking
out-of-equilibrium constraints on scatterings involving heavy Majorana
neutrinos into account \cite{MAL}, one finds that $\varepsilon'$ may
reach values up to $10^{-7} - 10^{-6}$ in such a scenario. This must
be compared with the usual scenario in \cite{epsilonprime}, for which
$\varepsilon'<10^{-15}$ for $m_{Ni}\approx 1$ TeV, and hence very
heavy neutrinos are needed to account for the BAU.

In Refs.\ \cite{IKS,KRS}, the authors pointed out that CP
violation in heavy particle decays responsible for the baryon (lepton)
asymmetry may be further enhanced if one considers the absorptive part
of the Higgs self-energies, which was though neglected in subsequent
studies.  Since this kind of CP violation may resemble the known
mechanism of CP violation through $K^0\bar{K}^0$ mixing in the kaon
complex \cite{reviewCP}, we shall call it hereafter as CP violation of
the $\varepsilon$ type.  Exploiting this idea, Botella and Roldan
\cite{BR} gave some estimates for $\varepsilon$-type CP violation in
Higgs decays.  They found that the ratio $\varepsilon/\varepsilon'$
may indeed be large in an extended SU(5) unified model, since the
self-energy and vertex contributions to CP violation may have
different Yukawa coupling structures.  Applying this mechanism, the
authors in \cite{LS1} concluded that $\varepsilon/\varepsilon'$ is of
order unity in scenarios for leptogenesis, and hence the known results
\cite{FY} obtained for $\varepsilon'$-type CP violation may not need
be modified drastically.  Moreover, using an effective Hamiltonian
approach based on the Weisskopf-Wigner approximation (WW) \cite{WW},
the same authors \cite{LS2} reached the conclusion that
$\varepsilon$-type CP violation may be rather suppressed, when the two
mixed heavy Majorana neutrinos are nearly degenerate, and CP violation
vanishes completely in the limit, in which the two mass eigenvalues of
the effective Hamiltonian are {\em exactly} equal.

Recently, there has been renewed interest in the $\varepsilon$-type CP
violation due to the mixing of heavy Majorana neutrinos and the
implications of this mechanism for the BAU \cite{Paschos}. It has been
observed in \cite{Paschos} that CP violation can be considerably
enhanced through the mixing of two nearly degenerate heavy Majorana
neutrinos.  Using exact solutions for the wave functions, which were
obtained from diagonalizing the effective Hamiltonian, the authors
\cite{Paschos} have calculated $\varepsilon$ and found that it can be
larger than $\varepsilon'$ by two or even three orders of magnitude.
This result makes the leptogenesis scenario very attractive.  The
enhancement of the CP-violating phenomenon is in agreement with
earlier articles on resonant CP violation in scatterings involving top
quarks, supersymmetric quarks or Higgs particles in the intermediate
state \cite{APCP,APCP',APRL,AP}, as well as with a remark \cite{KRS}
concerning CP violation in the decays of exotic neutral leptons to
quarks.

The existing difference between earlier articles, which found values
of $\varepsilon/\varepsilon'$ of order one \cite{LS1,LS2}, and recent
authors \cite{APRL,Paschos,AP}, who discovered that $\varepsilon$
could be even of order unity \cite{APRL,AP}, may be attributed to the
problem of the proper treatment of two nearly degenerate states. It is
known that conventional perturbation field theory breaks down in the
limit of degenerate particles. For example, the wave-function
amplitude that describes the CP-asymmetric mixing of two heavy
Majorana neutrinos, $N_1$ and $N_2$, say, is inverse proportional to
the mass splitting $m_{N_1}-m_{N_2}$, and it becomes singular if the
degeneracy is exact \cite{LS2}. Solutions to this problem have been
based on the wave-function formalism in the WW approximation
\cite{LS2,Paschos}. Obviously, a more rigorous field-theoretic
approach to the resonant phenomenon of CP violation through nearly
degenerate heavy Majorana neutrinos is still necessary.  Therefore, it
is rather important to provide a field-theoretic solution to the
problem of $\varepsilon$-type CP violation and compare the so-derived
results with those found with other methods.

Since the dynamics of $\varepsilon$-type CP violation is quite closely
related with CP violation induced by particle widths \cite{APCP}, one
is therefore compelled to rely on resummation approaches, which treat
unstable particles in a consistent way.  In the context of gauge field
theories, a gauge-independent resummation approach to resonant
transition amplitudes has been formulated, which is implemented by the
pinch technique \cite{PP}. Subsequently, this formalism has been
extended to the case of mixing between two intermediate resonant
states in scattering processes \cite{APRL,AP}. Here, we develop a
related formalism for decays, which can effectively take into account
phenomena of mixing of states during the decay of particles.

Consequently, our main interest in this paper will be to study the
$\varepsilon$- and $\varepsilon'$-type mechanisms of CP violation in
some detail, within the framework of an effective field-theoretic
formalism devised for decay amplitudes.  This formalism consistently
describes the phenomenon of resonantly enhanced CP violation through
the mixing of nearly degenerate heavy neutrinos and can therefore be
applied to any analogous system responsible for baryogenesis, in which
CP violation is of the $\varepsilon$ type. The analytic results
obtained for $\varepsilon$-type CP violation with our field-theoretic
approach do not display singularities and exhibit a physically correct
analytic behaviour in transition amplitudes. The fact that resonant CP
violation through mixing may be of order one \cite{APRL,AP} can lead
to scenarios, in which the heavy Majorana neutrinos are relatively
light with masses as low as 1 TeV.  This is roughly the highest
mass scale, at which the electroweak phase transitions can still
occur.  Most interestingly, this mechanism can produce significant
$\varepsilon$-type CP violation, even if all Yukawa couplings are of
the same magnitude and the Majorana masses are of few TeV.  This novel
phenomenological consequence of resonant CP violation has not yet been
studied in this leptogenesis scenario. Furthermore, it is worth
investigating the influence of other possible phenomena on this resonant
CP-violating mechanism, such as low-energy constraints due to the
electric dipole moment (EDM) of electron or finite-temperature
effects.

The paper is organized as follows. In Section 2, we describe minimally
extended models that include heavy Majorana neutrinos.  At low
energies, these models amount to adding isosinglet neutrino states to
the field content of an effective one-Higgs doublet model. Such
scenarios may be embedded into certain SO(10) and/or E$_6$ unified
theories, which can naturally predict nearly degenerate heavy Majorana
neutrinos as light as 100 GeV. Moreover, we discuss the
renormalizability of the effective model.  In Section 3, we present a
resummation formalism for resonant transitions between fermions in
decay amplitudes. Furthermore, we illustrate some of the advantages of
this approach, when compared to other existing methods.  Making use of
our formalism, we calculate the analytic expressions of the relevant
transition amplitudes in Section 4. In Section 5, we give estimates of
possible constraints coming from low-energy data, such as the EDM of
electron, which turn out to be quite weak in order to rule out our
leptogenesis scenario.  In Section 6, we present the Boltzmann
equations relevant for the evolution of the leptonic asymmetry in the 
effective model, and give numerical estimates and comparisons
for the BAU generated via $\varepsilon$- and/or $\varepsilon'$-type CP
violation.  We also discuss the implications of finite temperature
effects for the resonant phenomenon of CP violation and find that such
a phenomenon can still be viable.  We draw our conclusions in Section 7.

\setcounter{equation}{0}
\section{Heavy Majorana neutrino models}

Heavy Majorana neutrinos may naturally be realized in certain GUT's,
such as SO(10) \cite{FM,Wol/Wyl} and/or E${}_6$ \cite{witten} models.
Nevertheless, these models will also predict several other particles,
{\em e.g.}, leptoquarks, additional charged and neutral gauge bosons
($W^\pm_R$ and $Z_R$), which may deplete the number density of heavy
neutrinos $N_i$ through processes of the type $N_i\bar{e}_R \to
W^{+*}\to u_R d_R$ and so render the whole analysis very involved.  If
these particles are sufficiently heavier than the lightest heavy
Majorana neutrino and/or the temperature of the universe \cite{MAL},
this problem may be completely avoided. Since we wish to simplify our
analysis without sacrificing any of the essential features involved in
the study of the BAU, we shall consider a minimal model with
isosinglet neutrinos, which is invariant under the SM gauge group,
SU$(2)_L\otimes$U$(1)_Y$. Then, we will present the relevant
Lagrangians that govern the interactions of the heavy Majorana
neutrinos with the Higgs fields and the ordinary leptons. Also, we
will identify the non-trivial CP-violating phases of the model and pay
special attention to the one-loop renormalization of the Yukawa
couplings.

As has been mentioned above, certain SO(10) \cite{FM,Wol/Wyl} and/or
E${}_6$ \cite{witten} models naturally predict the existence of heavy
Majorana neutrinos. In SO(10) models, an attractive breaking pattern
down to the SM may be given schematically in the following way:
\begin{eqnarray}
\mbox{SO(10)} &\to & G_{422}=\mbox{SU}(4)_{\mbox{\scriptsize PS}}
                              \otimes\mbox{SU}(2)_R \otimes 
                                                     \mbox{SU}(2)_L\nonumber\\
&\to & G_{3221} =
\mbox{SU}(3)_c\otimes \mbox{SU}(2)_R\otimes \mbox{SU}(2)_L\otimes 
                                          \mbox{U}(1)_{(B-L)}\nonumber\\
&\to & \mbox{SM} = G_{321} = \mbox{SU}(3)\otimes \mbox{SU}(2)_L\otimes 
\mbox{U}(1)_Y \, , 
\end{eqnarray}
where the subscript PS refers to the Pati-Salam gauge group \cite{PS}.
The spinor representation of SO(10) is 16 dimensional and its
decomposition under $G_{422}$ is given by
\begin{equation}
\label{G422}
G_{422}\, :\ \mbox{{\bf 16}}\ \to\ (4,1,2)\, \oplus\, (\bar{4},2,1)\, .
\end{equation}
As can be seen from Eq.\ (\ref{G422}), it is evident that SO(10) can
accommodate right-handed neutrinos, since it contains the left-right
symmetric gauge group SU(2)$_R\otimes$SU(2)$_L\otimes$U(1)$_{(B-L)}$.
There are several Higgs-boson representation that can give rise to the
breakdown of $G_{422}$ and $G_{3221}$ down to the SM gauge group
$G_{321}$ \cite{Wol/Wyl,PalMoh}. In E$_6$ models \cite{witten}, the
{\bf 27} spinor representation decomposes into {\bf 16} $\oplus$ 
{\bf 10} $\oplus$ {\bf 1} under SO(10), which leads to four singlet
neutrinos per SM family: one neutrino as isodoublet member in {\bf 16}, 
two neutrinos as isodoublet members in {\bf 10}, and one singlet
neutrino in {\bf 1}. In these models, two of the four isosinglets can
have Majorana masses of few TeV \cite{witten}, depending on the
representation of the E${}_6$ Higgs multiplets, whereas the other two
are very heavy with masses of the order of the unification scale.
Possible flavour structures for the isosinglet neutrino mass matrix
resulting from the above two representative unified models will be 
discussed below.

The minimal model under consideration extends the SM field content of
the three lepton and quark families by adding a number $n_R$
right-handed neutrinos $\nu_{Ri}$, with $i=1, 2, \dots, n_R$.  Even
though in E$_6$ models the active isosinglet neutrinos may be more
than three, in the SO(10) models mentioned above the symmetric case of
having one right-handed neutrino per family turns out to be quite
natural, {\em i.e.}, $n_R=3$. Therefore, we shall not specify the
number of $\nu_{Ri}$ in the following. To be specific, the leptonic
sector of our minimal model consists of the fields:
\begin{displaymath}
\left( \begin{array}{c} \nu_{lL} \\ l_L \end{array} \right)\ ,\qquad l_R\ ,
\qquad \nu_{Ri}\ ,
\end{displaymath}
with $l=e,\mu,\tau$. Since for temperatures $T\gg T_c
\stackrel{\displaystyle >}{\sim} v$, one has $v(T)=0$, where $v(T)$ is the
vacuum expectation value (VEV) of the SM Higgs doublet $\Phi$ at temperature
$T$ (with $v=v(0)$), the only admissible mass terms are those of the Majorana
type and are given by the Lagrangian
\begin{equation}
\label{Majmass}
-{\cal L}_M\ =\ \frac{1}{2}\, \sum\limits_{i,j=1}^{n_R}\, 
\Big( \bar{\nu}^C_{Ri} M^\nu_{ij} \nu_{Rj}\ +\ 
\bar{\nu}_{Ri} M^{\nu *}_{ij} \nu^C_{Rj}\, \Big)\, .
\end{equation}
Here, the superscript $C$ denotes the operation of charge conjugation,
which acts on the four-component chiral spinors $\psi_L$ and $\psi_R$
as $(\psi_L)^C=P_R C\bar{\psi}^T$ and $(\psi_R)^C=P_L C\bar{\psi}^T$,
where $P_{L(R)}=[1-(+)\gamma_5]/2$ is the chirality projection
operator.  In Eq.\ (\ref{Majmass}), $M^\nu$ is a $n_R\times n_R$
dimensional symmetric matrix, which is in general complex. The
isosinglet mass matrix $M^\nu$ can be diagonalized by the unitary
transformation $U^T M^\nu U = \widehat{M}^\nu$, where $U$ is a
$n_R\times n_R$ dimensional unitary matrix and $\widehat{M}^\nu$ is a
positive diagonal matrix containing the $n_R$ heavy Majorana masses.
Furthermore, the $n_R$ mass eigenstates $N_i$ are related to the
flavour states $\nu_{Ri}$ through $\nu_{Ri} = P_R\sum_{j=1}^{n_R}
U_{ij}N_j$. In the basis, in which the isosinglet neutrino mass matrix
is diagonal and equals $\widehat{M}^\nu$, the Yukawa sector describing
the interaction of the heavy neutrinos with the Higgs doublet and the
ordinary leptons reads
\begin{equation}
\label{LYint}
{\cal L}_Y\ =\ -\, \sum\limits_{l=1}^{n_L} \sum\limits_{j=1}^{n_R}\,
h_{lj}\, (\bar{\nu}_{lL},\ \bar{l}_L)\, 
\left( \begin{array}{c} (H\, -\, i\chi^0)/\sqrt{2} \\ -\, \chi^- \end{array}
\right)\, N_j\ +\ \mbox{H.c.}
\end{equation}
The CP-even Higgs field $H$, the CP-odd Higgs scalar $\chi^0$ and the
charged Higgs scalars $\chi^\pm$ given in (\ref{LYint}) are all massless
at high temperatures. In the limit $T\to 0$, $H$ represents the massive SM
Higgs boson, whereas $\chi^0$ and $\chi^\pm$ are the massless would-be
Goldstone bosons eaten by the longitudinal degrees of freedom of the gauge
bosons $Z$ and $W^\pm$, respectively.  If the SM Higgs field $H$ is very
heavy with mass of order 1 TeV at $T=0$ and also close to the mass of the
decaying heavy neutrinos, then mass Higgs effects may not be negligible.
Therefore, we shall initially keep the full $M_H$ dependence in our
calculations and then present analytic results for the limiting case
$M_H=0$.

Before counting all the non-trivial CP-violating phases for the most
general case, it may be more instructive to discuss first some simple
models that could predict degenerate or almost degenerate heavy
Majorana neutrinos and CP violation. To this end, we consider a
minimal scenario, in which the fermionic matter of the SM is extended
by adding two right-handed neutrinos per family, {\em e.g.},
$\nu_{lR}$ and $(S_{lL})^C$ with $l=e,\mu,\tau$. Such a scenario may
be derived from certain SO(10) \cite{Wol/Wyl} and/or E$_6$
\cite{witten} models. For our illustrations, we neglect possible
inter-family mixings. Imposing lepton-number conservation on the model
gives rise to the Lagrangian
\begin{equation}
\label{WW}
- {\cal L}\ =\ \frac{1}{2}\, (\bar{S}_L,\ (\bar{\nu}_R)^C )\, 
\left( \begin{array}{cc}
0 & M \\
M & 0 \end{array} \right)\, \left( \begin{array}{c}
(S_L)^C \\ \nu_R \end{array} \right)\ +\ 
h_R\, (\bar{\nu}_L,\ \bar{l}_L) \tilde{\Phi} \nu_R\ +\ \mbox{H.c.},
\end{equation}
where $\tilde{\Phi}=i\sigma_2\Phi$ is the isospin conjugate Higgs doublet
and $\sigma_2$ is the usual Pauli matrix. Even though the mass and coupling 
parameters may be complex in such a scenario, the phase redefinitions of 
the fields,
\begin{equation}
\label{CProt}
\nu_L \to e^{i\phi_\nu} \nu_L,\qquad 
l_L \to e^{i\phi_l} l_L,\qquad
\nu_R \to e^{i\phi_R} \nu_R,\qquad
S_L \to e^{i\phi_L} S_L ,
\end{equation}
can, however, make them all real.  This model is CP invariant, unless one
allows for a non-trivial mixing among generations \cite{BRV}. Moreover,
the model preserves the lepton number and hence cannot produce any excess
in $L$ through heavy neutrino decays.

There are two equivalent ways to break the $L$ and CP invariance of
the Lagrangian in Eq.\ (\ref{WW}): One has to either (i) introduce two
complex $L$-violating mass terms of the kind $\mu_R\bar{\nu}_R\nu^C_R$
and $\mu_L\bar{S}^C_L S_L$, where both $\mu_R$ and $\mu_L$ are
complex, or equivalently (ii) add the $L$-violating coupling 
$h_R\, (\bar{\nu}_L,\ \bar{l}_L) \tilde{\Phi} (S_L)^C$ and include
the $L$-violating mass parameter, {\em e.g.}, $\mu_R\bar{\nu}_R\nu^C_R$.
These are the minimal enlargements that can assure $L$ and CP violation 
on the same footing in this simple two-isosinglet neutrino model. 
In fact, the necessary conditions for CP invariance in these two scenarios 
are written down
\begin{eqnarray}
\label{CPi_ii}
\mbox{(i)}  && |h_R|^2\, \Im m ( M^{*2}\mu_L\mu_R )\ =\ 0\, ,\nonumber\\
\mbox{(ii)} &&  \Im m (h_Lh_R^*\mu_R M^*)\ =\ 0\, . 
\end{eqnarray}
It can easily be checked that the two equalities in Eq.\ 
(\ref{CPi_ii}) are invariant under the phase redefinitions of the
fields given in Eq.\ (\ref{CProt}). We must remark that the
$L$-breaking parameters $\mu_L$ and $\mu_R$ are generally much smaller
than $M$ within E$_6$ scenarios. The origin of these parameters are
usually due to residual effects of high-dimensional operators
involving super-heavy neutrinos \cite{witten}.  The typical size of
the $L$-violating parameters is $\mu_L,\ \mu_R \sim M^2/M_{Planck}$,
$M^2/M_X$ or $M^2/M_S$, where $M_S\approx 10^{-3}\, M_X$ is some
intermediate see-saw scale. As a consequence, such effective minimal
models derived from E$_6$ theories can naturally predict small mass
splittings for the heavy neutrinos $N_1$ and $N_2$. To a good
approximation, this small mass difference may be determined from the
parameter $x_N=m_{N_2}/m_{N_1}-1\sim \mu_L/M$ or $\mu_R/M$.  For
instance, if $M=10$ TeV and $\mu_L=\mu_R = M^2/M_X$, one then finds
$x_N\approx 10^{-12} - 10^{-11}$. As we will see in Section 4, these
small values for the mass difference $x_N$ can produce large CP
asymmetries in the heavy neutrino decays.

In order to deduce the sufficient and necessary conditions for the
most general structure of the two right-handed neutrino model, one
must consider the systematic approach presented first in \cite{BBG},
which is slightly different from the procedure outlined above. In this
approach \cite{BRV,BBG}, one looks for all possible weak-basis
independent combinations that can be formed by Yukawa couplings and
the neutrino mass matrix $M^\nu$, and are simultaneously invariant
under generalized CP transformations of the fields. These generalized
CP transformations of the fermion fields may include unitary flavour
rotations, apart from the phase redefinitions mentioned above. Further
details may be found in Ref.\ \cite{BRV}. Thus, for the model at hand,
we find that the sufficient and necessary condition for CP invariance
is
\begin{equation}
\label{CPinv}
\Im m\, \mbox{Tr} ( h^\dagger h M^{\nu\dagger} M^\nu M^{\nu\dagger}
h^T h^* M^\nu )\ =\ m_{N_1}m_{N_2} (m^2_{N_1} - m^2_{N_2})\,
\Im m (h_{l1}h^*_{l2})^2\ =\ 0\, .
\end{equation}
The $1\times 2$ dimensional matrix $h$ in Eq.\ (\ref{CPinv}) contains
the Higgs Yukawa couplings, which are defined as $h_{lj}$ in Eq.\ 
(\ref{LYint}), {\em i.e.}, in the physical mass basis where
$M^\nu=\widehat{M}^\nu$. One can show that Eq.\ (\ref{CPinv}) is
consistent with the conditions in Eq.\ (\ref{CPi_ii}) for the two
special cases discussed above. From Eq.\ (\ref{CPinv}), one readily
sees that only one physical CP-violating combination is possible in
this minimal model and CP invariance is restored if $m_{N_1}=m_{N_2}$
provided none of the isosinglet neutrinos is massless.  The above
considerations may be extended to models with more than two
right-handed neutrinos and more than one lepton families. In this
case, there may be more conditions analogous to Eq.\ (\ref{CPinv}),
which involve high order terms in the Yukawa-coupling matrix $h$.
However, not all of the conditions are sufficient and necessary for CP
invariance.  Instead of undertaking the rather difficult task to
derive all CP-invariant conditions, we note in passing that the total
number ${\cal N}_{CP}$ of all non-trivial CP-violating phases in a
model with $n_L$ weak isodoublets and $n_R$ neutral isosinglets
is ${\cal N}_{CP} = n_L(n_R-1)$ \cite{KPS}.

%******************************************************************
%%%Figure 1
%******************************************************************
\begin{center}
\begin{picture}(360,300)(0,0)
\SetWidth{0.8}

\ArrowLine(0,270)(30,270)\ArrowLine(30,270)(60,270)\Line(60,270)(60,230)
\ArrowLine(90,230)(60,230)\DashArrowLine(60,270)(90,290){5}
\DashArrowLine(30,270)(60,230){5}\Text(60,250)[]{{\boldmath $\times$}}
\Text(0,265)[lt]{$N_i$}\Text(45,280)[]{$l$}\Text(65,250)[l]{$N_j$}
\Text(95,290)[l]{$\chi^-$}\Text(95,230)[l]{$l$}\Text(40,250)[r]{$\chi^+$}
\Text(50,210)[]{\bf (a)}

\ArrowLine(120,270)(150,270)\ArrowLine(150,270)(180,270)
\ArrowLine(180,270)(180,230)\ArrowLine(180,230)(210,230) 
\DashArrowLine(180,270)(210,290){5}\DashArrowLine(150,270)(180,230){5}
\Text(120,265)[lt]{$N_i$}\Text(165,280)[]{$\nu_l$}\Text(185,250)[l]{$N_j$}
\Text(215,290)[l]{$\chi^0$}\Text(215,230)[l]{$\nu_l$}
\Text(170,240)[r]{$\chi^0,H$}
\Text(170,210)[]{\bf (b)}

\ArrowLine(240,270)(270,270)\ArrowLine(270,270)(300,270)
\ArrowLine(300,270)(300,230)\ArrowLine(300,230)(330,230) 
\DashArrowLine(300,270)(330,290){5}\DashArrowLine(270,270)(300,230){5}
\Text(240,265)[lt]{$N_i$}\Text(285,280)[]{$\nu_l$}\Text(305,250)[l]{$N_j$}
\Text(335,290)[l]{$H$}\Text(335,230)[l]{$\nu_l$}
\Text(290,240)[r]{$\chi^0,H$}
\Text(290,210)[]{\bf (c)}

\DashArrowLine(0,150)(30,150){5}\ArrowArc(50,150)(20,0,180)
\ArrowArc(50,150)(20,180,360)\DashArrowLine(70,150)(100,150){5}
\Text(0,155)[bl]{$\chi^-$} \Text(100,155)[br]{$\chi^-$}
\Text(50,175)[b]{$N_i$}\Text(50,125)[t]{$l$}
\Text(50,100)[]{\bf (d)}

\DashArrowLine(120,150)(150,150){5}\ArrowArc(170,150)(20,0,180)
\ArrowArc(170,150)(20,180,360)\DashArrowLine(190,150)(220,150){5}
\Text(120,155)[bl]{$\chi^0$} \Text(220,155)[br]{$\chi^0$}
\Text(170,175)[b]{$N_i$}\Text(170,125)[t]{$\nu_l$}
\Text(170,100)[]{\bf (e)}

\DashArrowLine(240,150)(270,150){5}\ArrowArc(290,150)(20,0,180)
\ArrowArc(290,150)(20,180,360)\DashArrowLine(310,150)(340,150){5}
\Text(240,155)[bl]{$H$} \Text(340,155)[br]{$H$}
\Text(290,175)[b]{$N_i$}\Text(290,125)[t]{$\nu_l$}
\Text(290,100)[]{\bf (f)}

\ArrowLine(0,40)(30,40)\ArrowLine(30,40)(70,40)
\ArrowLine(70,40)(100,40)\DashArrowArc(50,40)(20,0,180){5}
\Text(0,45)[bl]{$l'$}\Text(100,45)[br]{$l$}
\Text(50,65)[b]{$\chi^+$}\Text(50,35)[t]{$N_i$}
\Text(50,0)[]{\bf (g)}

\ArrowLine(120,40)(150,40)\ArrowLine(150,40)(190,40)
\ArrowLine(190,40)(220,40)\DashArrowArc(170,40)(20,0,180){5}
\Text(120,45)[bl]{$\nu_{l'}$}\Text(220,45)[br]{$\nu_l$}
\Text(170,65)[b]{$\chi^0,H$}\Text(170,35)[t]{$N_i$}
\Text(170,0)[]{\bf (h)}

\ArrowLine(240,40)(270,40)\ArrowLine(270,40)(310,40)
\ArrowLine(310,40)(340,40)\DashArrowArc(290,40)(20,0,180){5}
\Text(240,45)[bl]{$N_j$}\Text(340,45)[br]{$N_i$}
\Text(290,65)[b]{$\chi^\mp,\chi^0,H$}\Text(290,35)[t]{$l^\mp,N_i,N_i$}
\Text(290,0)[]{\bf (j)}

\end{picture}\\[0.7cm]
\end{center}
\begin{list}{}{\labelwidth1.6cm\leftmargin2.5cm\labelsep0.4cm\itemsep0ex 
plus0.2ex }

\item[{\small {\bf Fig.\ 1:}}] 
{\small One-loop graphs contributing to the renormalization of the 
couplings $\chi^-lN_i$, $\chi^0\nu_l N_i$ and $H\nu_lN_i$.}
\end{list}

Since CP violation in $N_i$ decays will necessitate non-zero one-loop
absorptive parts of vertex and $N_i$ self-energy graphs as will be
seen in Section 4, one should also have to address the issue of
renormalization of the dispersive counterparts (see also Fig.\ 1).  In
order to check explicitly that our minimal model leads indeed to
consistent renormalizable results, we shall adopt the following
strategy in our analysis. First, we fix the renormalization of all
Higgs Yukawa couplings $h_{lj}$ from the decay mode $N_i\to
l^+\chi^-$. In this way, we determine the counter-terms (CT) of
$h_{lj}$, $\delta h_{lj}$.  Then, we show that all ultra-violet (UV)
divergences cancel in the partial decays $N_i\to \nu_l\chi^0$ and
$N_i\to \nu_l H$.  For this purpose, we first express all bare
quantities in terms of renormalized ones as follows:
\begin{eqnarray}
\label{Rbare}
\nu^0_{lL} &=& \sum\limits_{l'=1}^{n_L}\, \Big( \delta_{ll'}\, +\, 
\frac{1}{2}\delta Z^\nu_{ll'} \Big) \nu_{l'L}\, ,\qquad
l^0_L\ =\ \sum\limits_{l'=1}^{n_L}\, \Big( \delta_{ll'}\, +\, 
\frac{1}{2}\delta Z^l_{ll'} \Big) l'_L\, ,\\
N^0_i &=& \sum\limits_{j=1}^{n_R}\, \Big( \delta_{ij}\, +\, 
\frac{1}{2}\delta Z^N_{ij} \Big) N_j\, ,\qquad
\tilde{\Phi} \ =\ \Big( 1\, +\, \frac{1}{2}\delta Z_\Phi \Big) \tilde{\Phi}
\, ,\qquad
h^0_{lj} \ =\ h_{lj}\ +\ \delta h_{lj}\, .\nonumber 
\end{eqnarray}
The superscript `0' in Eq.\ (\ref{Rbare}) indicates that the field or
coupling parameter is unrenormalized, whereas quantities without this
superscript are considered to be renormalized. In addition, the CT
$\delta Z_\Phi$ collectively denotes the wave-function renormalization
constants of all components of the Higgs doublet $\tilde{\Phi}$ (or
$\Phi$), {\em i.e.}, the fields $\chi^\pm$, $\chi^0$ and $H$. The
divergent part of all the Higgs wave-function renormalizations,
$\delta Z^{div}_\Phi$, has been found to be universal.  Expressions
showing the universality of $\delta Z^{div}_\Phi$ together with other
relevant one-loop analytic results are relegated to Appendix A. Taking
the relations in Eq.\ (\ref{Rbare}) into account, we find that the
renormalized Lagrangian in (\ref{LYint}) gets shifted by an amount
\begin{equation}
\label{deltaLY}
-\, \delta{\cal L}_Y\  =\ \frac{1}{2}\, \sum\limits_{l=1}^{n_L}
\sum\limits_{j=1}^{n_R}\, \Big(\, 2\frac{\delta h_{lj}}{h_{lj}}\, +\,
\delta Z_\Phi\, +\, \sum\limits_{l'=1}^{n_L} \delta Z^L_{l'l}\, +\,
 \sum\limits_{k=1}^{n_R} \delta Z^N_{jk}\, \Big)
\, \bar{L}_{l'}\tilde{\Phi} N_k\ +\ \mbox{H.c.},  
\end{equation}
where $L_l=(\nu_{lL},\ l_L )^T$ and $\delta Z^L = (\delta Z^l, \delta
Z^\nu)$. From Fig.\ 1(a), it is easy to see that the one-loop
correction to the coupling $\chi^+ Nl$ can only occur via a $\Delta
L=2$ Majorana mass insertion, owing to charge conservation on the
vertices. Naive power counting may then convince oneself that the
one-loop irreducible vertex $\chi^+ Nl$ is UV finite. Less obvious is
the UV finiteness for the proper couplings $\chi^0N\nu$ and $HN\nu$,
which is shown in Appendix A.

As has been discussed above, we shall now determine the Yukawa
coupling CT's $\delta h_{lj}$ from the renormalization of the coupling
$\chi^+ Nl$. Requiring that all UV terms are absorbed into the 
definition of $h_{lj}$, we obtain 
\begin{equation}
\label{deltah}
\delta h_{lj}\ =\ -\, \frac{1}{2}\, \Big(\, h_{lj}\delta Z_{\chi^-}\,
+\, \sum\limits_{l'=1}^{n_L} h_{l'j}\delta Z^{l*}_{l'l}\, +\,
\sum\limits_{k=1}^{n_R} h_{lk}\delta Z^N_{kj}\, \Big)\, .
\end{equation}
We observe that $\delta h_{lj}$ may be separated into two terms: The
wave-function term $\delta Z_{\chi^-}$, which is flavour independent,
and the rest, which depends on the flavour and the wave-function CT's
$\delta Z^{l}_{ll'}$ and $\delta Z^N_{ij}$.  If we had renormalized
the Higgs Yukawa couplings from the decays $N\to \nu H$, the only
difference in Eq.\ (\ref{deltah}) would have been the appearance of
the CT's $\delta Z_H$ and $\delta Z^\nu_{ll'}$ in place of $\delta
Z_{\chi^-}$ and $\delta Z^{l}_{ll'}$, respectively. Also, for the
decay $N\to \nu\chi^0$, one has to include the wave-function
renormalization of $\chi^0$, $\delta Z_{\chi^0}$, instead of $\delta
Z_H$ in the decay $N\to \nu H$. Therefore, consistency of Yukawa
coupling renormalization requires that differences of the kind $\delta
Z^l_{ll'} - \delta Z^\nu_{ll'}$, $\delta Z_{\chi^0}-\delta Z_H$ and
$\delta Z_{\chi^-}-\delta Z_H$ must be UV safe. In Appendix A, it is
shown that all these CT differences are indeed UV finite and vanish in
the limit of $M_H\to 0$. This completes our discussion concerning the
one-loop renormalization of heavy Majorana neutrino decays. In the
next section, we shall explicitly demonstrate how the renormalization
presented here gets implemented within our resummation formalism for
unstable particle mixing.

\setcounter{equation}{0}
\section{Resummation approach for two-fermion mixing}

If a Lagrangian contains unstable particles, then these fields cannot
be described by free plane waves at times $t\to \pm \infty$ and hence
cannot formally appear as asymptotic states in the conventional
perturbation field theory. Within a simple scalar theory with one
unstable particle, Veltman~\cite{Velt} showed that, even if one
removes the unstable particle from the initial and final states and
substitutes it in terms of asymptotic states, the so-truncated
S-matrix theory will still maintain the field-theoretic properties of
unitarity and causality. Our main concern in this section will be to
present an approach to decay amplitudes that describes the dynamics of
unstable particle mixing.  Hence, in such a formulation, finite width
effects in the mixing and decay of non-asymptotic states must be taken
into account.  This will be done in an effective manner, such that the
decay amplitude derived with this method can be embedded in an
equivalent form to a transition element \cite{PP,AP} in agreement with
Veltman's S-matrix approach.  This effective field-theoretic approach
is equivalent to that of the decay of an initial pure state, such as
the states $K^0$ or $\bar{K}^0$, which is initially produced by some
asymptotic states in kaon experiments, {\em e.g.}, in $p\pi^-$ or
$p\bar{p}$ collisions \cite{reviewCP}. Since the time evolution of the
decaying system is effectively integrated out over all times, the
resummed decay amplitudes derived with this field-theoretic method
will not display any explicit time dependence.

The discussion in this section is organized as follows.  First, we
briefly review the theoretical description of the mixing between
stable particles in a simple scalar theory within the framework of the
Lehmann--Symanzik--Zimmermann formalism (LSZ) \cite{LSZ}. After
gaining some insight, we extend our considerations to the mixing
between two unstable scalars. The effective field-theoretic method
developed for the scalar case can then carry over to the case of
mixing of two unstable fermions, with the help of which
$\varepsilon$-type CP violation will be calculated in Section 4.

Let us now consider a field theory with $N$ real scalars $S^0_i$, with
$i=1,2,\dots,N$. We shall assume that the scalars are stable to a good
approximation and neglect possible finite width effects of the 
particles. The bare (unrenormalized) fields $S^0_i$ and their respective
masses $M^0_i$ may then be expressed in terms of renormalized fields
$S_i$ and masses $M_i$ in the following way:
\begin{eqnarray}
\label{RenS0}
S^0_i & = & Z^{1/2}_{ij}\, S_j \ =\ \Big( \delta_{ij}\, +\,
\frac{1}{2} \delta Z_{ij}\Big) S_j\ ,\\
\label{RenMass}
(M^0_i)^2 & = & M^2_i\, +\, \delta M^2_i\ .
\end{eqnarray}
Here and in the following, summation is understood over repeated
indices that do not appear on both sides of an equation.  In Eqs.\ 
(\ref{RenS0}) and (\ref{RenMass}), $Z^{1/2}_{ij}$ and $\delta M_i$ are
the wave-function and mass renormalization constants, respectively,
which can be determined from renormalization conditions imposed on the
two-point correlation functions, $\Pi_{ij}(p^2)$, for the transitions
$S_j\to S_i$ in some physical scheme, such as the on-mass-shell (OS)
renormalization scheme \cite{OS}.

%******************************************************************
%%%Figure 2
%******************************************************************
\begin{center}
\begin{picture}(360,100)(0,0)
\SetWidth{0.8}

\Text(0,60)[l]{$S_{i,\dots}$}\Text(30,60)[]{$=$}
\Text(60,60)[]{$\lim\limits_{p^2\to M^2_i}$}
\GOval(150,60)(30,20)(0){0.5}\Line(170,60)(200,60)\GCirc(215,60){15}{0.5}
\Line(230,60)(260,60)\Vertex(260,60){2}
\Line(120,90)(135,80)\Line(120,30)(135,40)\Vertex(120,90){2}
\Vertex(120,30){2}\Vertex(113,60){2}\Vertex(115,75){2}\Vertex(115,45){2}
\Text(115,90)[r]{$S_{i_1}$}\Text(115,30)[r]{$S_{i_n}$}
\Text(185,65)[b]{$S_k$}\Text(245,65)[b]{$S_j$}
\Text(270,80)[l]{$Z^{-1/2T}_{ji}\, \hat{\Delta}^{-1}_{ii}(p^2)$}
\LongArrow(250,50)(235,50)\Text(255,50)[l]{$p$}

\end{picture}\\[0.7cm]
\end{center}
\begin{list}{}{\labelwidth1.6cm\leftmargin2.5cm\labelsep0.4cm\itemsep0ex 
plus0.2ex }

\item[{\small {\bf Fig.\ 2:}}] {\small Diagrammatic representation of the 
renormalized $n-1$-non-amputated amplitude, $S_{i,\dots}$, and the LSZ 
reduction formalism.}
\end{list}

It  will prove useful for the discussion that follows to give the 
relation of the pole parts between the unrenormalized scalar
propagators $\Delta_{ij}(p^2)$ and the renormalized ones 
$\hat{\Delta}_{ij}(p^2)$. The two pole parts are related through \cite{OS} 
\begin{equation}
\label{Delpole}
\Delta_{ij}(p^2)\Big|_{p^2\to M^2_i,M^2_j}\ =\ 
Z^{1/2}_{im}\, \frac{\delta_{mn}}{p^2-M^2_n}\, Z^{1/2T}_{nj}\, .
\end{equation}
Using the LSZ formalism shown schematically in Fig.\ 2, one 
can deduce the renormalized $n-1$-non-amputated amplitude, 
$S_{i,\dots}$, for a fixed given external line $i$, from 
the corresponding unrenormalized $n$-point Green function $G_{i\dots}$,
where $n$ is the total number of external lines. 
In this way, we have
\begin{eqnarray}
\label{LSZ1}
S_{i,\dots}& = & \lim\limits_{p^2\to M^2_i}\
G_{j\dots}  Z^{-1/2T}_{ji} (p^2-M^2_i)\nonumber\\
&=& \lim\limits_{p^2\to M^2_i}\ {\cal T}^{amp}_{k,\dots}  
Z^{1/2}_{km}\, \frac{\delta_{mn}}{p^2-M^2_n}\, Z^{1/2T}_{nj}
Z^{-1/2T}_{ji} (p^2-M^2_i)\nonumber\\
&=& \lim\limits_{p^2\to M^2_i}\ {\cal T}^{amp}_{k,\dots}
Z^{1/2}_{ki}\ ,
\end{eqnarray}
where ${\cal T}^{amp}_{k,\dots}$ denotes the amplitude amputated at
the $k$ external leg. Clearly, the LSZ reduction procedure outlined
above can be generalized to all external legs, thus leading to the
physical (renormalized) $S$-matrix element $S_{i_1\dots i_n}$, which
governs the transition amplitude of $n$ asymptotic states.

Let us now consider the mixing of two neutral unstable scalars
\cite{APRL,AP}, {\em e.g.}, $S_1$ and $S_2$. Since we are interested
in studying the width effects of these particles, we have first to
calculate all the $S_iS_j$ Green functions, with $i,j =1,2$.  After
summing up a geometric series of the self-energies $\Pi_{ij}(p^2)$, the full
propagators may be obtained by inverting the following inverse
propagator matrix:
\begin{equation}
\label{InvD12}
\Delta^{-1}_{ij} (p^2)\ =\ 
\left[
\begin{array}{cc}
p^2\, -\, (M^0_1)^2\, +\, \Pi_{11}(p^2) & \Pi_{12}(p^2)\\
\Pi_{21}(p^2) & p^2\, -\, (M^0_2)^2\, +\, \Pi_{22}(p^2)
\end{array} \right]\, .
\end{equation}
The result of inverting the matrix in Eq.\ (\ref{InvD12}) may be given by
\begin{eqnarray} 
\label{D11}
\Delta_{11}(p^2) &=& \left[ \, p^2\, -\, (M^0_1)^2
+\Pi_{11}(p^2)-\, \frac{\Pi^2_{12}(p^2)}{p^2-(M^0_2)^2+ 
\Pi_{22}(p^2)}\right]^{-1}\,
,\\
\label{D22}
\Delta_{22}(p^2) &=& \left[ \, p^2\, -\, (M^0_2)^2
+\Pi_{22}(p^2)-\, \frac{\Pi^2_{12}(p^2)}{p^2-(M^0_1)^2+ 
\Pi_{11}(p^2)}\right]^{-1}\,
,\\
\label{D12}
\Delta_{12}(p^2) &=& \Delta_{21}(p^2)\ =\
-\, \Pi_{12}(s) \Bigg[ \Big(p^2-(M^0_2)^2+\Pi_{22}(p^2)\Big)
\Big( p^2 - (M^0_1)^2 +\Pi_{11}(p^2)\Big)\nonumber\\
&&-\, \Pi^2_{12}(p^2)\, \Bigg]^{-1}\, .
\end{eqnarray}
where $\Pi_{12}(p^2)=\Pi_{21}(p^2)$. Moreover, we find the useful
factorization property for the off-diagonal ($i\not=j$) resummed
scalar propagators
\begin{equation}
\label{Drel}
\Delta_{ij}(p^2)\ =\ -\, \Delta_{ii}(p^2)\
\frac{\Pi_{ij}(p^2)}{p^2\, -\, (M^0_j)^2\, +\,
\Pi_{jj}(p^2)}\  =\ -\, \frac{\Pi_{ij}(p^2)}{p^2\, -\, (M^0_i)^2\, +\,
\Pi_{ii}(p^2)}\ \Delta_{jj}(p^2)\ .
\end{equation} 
The resummed unrenormalized scalar propagators $\Delta_{ij}(p^2)$ are
related to the respective renormalized ones $\hat{\Delta}_{ij}(p^2)$
through the expression
\begin{equation} 
\label{D_Dhat}
\Delta_{ij}(p^2)\ =\ 
Z^{1/2}_{im}\, \hat{\Delta}_{mn}(p^2)\, Z^{1/2T}_{nj}\, ,
\end{equation}
where $\hat{\Delta}_{ij}(p^2)$ may be obtained from Eqs.\ 
(\ref{D11})--(\ref{D12}), just by replacing $M^0_i$ with $M_i$ and
$\Pi_{ij}(p^2)$ with $\widehat{\Pi}_{ij}(p^2)$.  Note that the
property given in Eq.\ (\ref{Drel}) will also hold true for the
renormalized scalar propagators $\hat{\Delta}_{ij}(p^2)$.  Taking
expressions (\ref{Drel}) and (\ref{D_Dhat}) into account, we can
derive the resummed and renormalized transition amplitude, denoted
here as $\widehat{S}_{i\dots}$, for the external leg $i$ which now
represents an unstable particle. This can be accomplished in
a way analogous to Eq.\ (\ref{LSZ1}), {\em viz.}
\begin{eqnarray}
\label{LSZ2}
\widehat{S}_{i\dots}& = & \lim\limits_{p^2\to M^2_i}\ 
{\cal T}^{amp}_{k\dots}  
Z^{1/2}_{km}\, \hat{\Delta}_{mn}(p^2)\, Z^{1/2T}_{nj}
Z^{-1/2T}_{ji} \hat{\Delta}^{-1}_{ii}(p^2) \nonumber\\
&=& \lim\limits_{p^2\to M^2_i}\Big[ {\cal T}^{amp}_{k\dots}Z^{1/2}_{ki}\ 
-\ {\cal T}^{amp}_{k\dots}Z^{1/2}_{km} \frac{\widehat{\Pi}_{mi}(p^2) 
(1-\delta_{mi})}{p^2-M^2_m+\widehat{\Pi}_{mm}(p^2)}\, \Big]\nonumber\\
&=& S_{i\dots}\ -\ S_{j\dots}\frac{\widehat{\Pi}_{ji}(M^2_i) 
(1-\delta_{ij})}{M^2_i-M^2_j+\widehat{\Pi}_{jj}(M^2_i)}\  ,
\end{eqnarray}
where $S_{i\dots}$ and $S_{j\dots}$ are the renormalized transition
elements evaluated from Eq.\ (\ref{LSZ1}) in the stable-particle
approximation. One should bear in mind that the OS renormalized
self-energies $\widehat{\Pi}_{ji}(M^2_i)$ in Eq.\ (\ref{LSZ2}) have no
vanishing absorptive parts, as renormalization can only modify the
dispersive (real) part of these self-energies. The reason is that the
CT Lagrangian must be Hermitian as opposed to the absorptive parts
which are anti-Hermitian. In fact, these additional width mixing
effects are those which we wish to include in our formalism for decay
amplitudes and are absent in the conventional perturbation theory. It
is also important to observe that our approach to decays is not
singular, {\em i.e.}, $\widehat{S}_{i\dots}$ displays an analytic
behaviour in the degenerate limit $M^2_i\to M^2_j$, because of the
appearance of the imaginary term $i\Im m\widehat{\Pi}_{jj}(M^2_i)$ in
the denominator of the mixing factor present in the last equality of
Eq.\ (\ref{LSZ2}).  Finally, we must stress that the inclusion of
these phenomena has been performed in an effective manner. Since the
decaying unstable particle cannot appear in the initial state
\cite{Velt}, the resummed decay amplitude must be regarded as being a
part which can effectively be embedded into a resummed $S$-matrix element
\cite{PP}.  This resummed $S$-matrix element describes the dynamics of
the very same unstable particle, which is produced by some asymptotic
states, resides in the intermediate state, and subsequently decays
either directly or indirectly, through mixing, into the observed final
states.

It is now straightforward to extend our considerations to the case of
mixing between two unstable fermions. Following a line of arguments
similar to those presented above, we consider a system with two
unstable fermions, call them $f_1$ and $f_2$. As usual, we express the
bare left- and right-handed chiral fields, $f^0_{Li}$ and $f^0_{Ri}$
(with $i=1,2$), in terms of renormalized fields as follows:
\begin{equation}
f^0_{Li}\ =\ Z^{1/2}_{Lij}\, f_{Lj}\ , \quad\qquad
f^0_{Ri}\ =\ Z^{1/2}_{Rij}\, f_{Rj}\ ,
\end{equation}
where $Z^{1/2}_{Lij}$ ($Z^{1/2}_{Rij}$) is the wave-function
renormalization constant for the left- (right-) handed chiral fields,
which may be determined from the fermionic self-energy transitions
$f_j\to f_i$, $\Sigma_{ij} (\not\!\! p)$, {\em e.g.}, in the OS
renormalization scheme \cite{KP}. Analogously with Eq.\ (\ref{InvD12}),
the resummed fermion propagator matrix may be obtained from
\begin{equation}
\label{InvS}
S_{ij}(\not\! p)\ =\ \left[ \begin{array}{cc} \not\! p - m^0_1 +
\Sigma_{11}(\not\! p) & \Sigma_{12}(\not\! p)\\
\Sigma_{21}(\not\! p) & \not\! p - m^0_2 + \Sigma_{22}(\not\! p)
\end{array} \right]^{-1},\qquad
\end{equation}
where $m^0_{1,2}$ are the bare fermion masses, which can be decomposed
into the OS renormalized masses $m_{1,2}$ and the CT mass terms
$\delta m_{1,2}$ as $m^0_{1,2}=m_{1,2} + \delta m_{1,2}$. Inverting
the matrix-valued $2\times 2$ matrix in Eq.\ (\ref{InvS}) yields
\begin{eqnarray}
\label{S11}
S_{11}(\not\! p) &=& \Big[\not\! p\, -\, m^0_1\, +\,
\Sigma_{11}(\not\! p)\, -\, \Sigma_{12}(\not\! p)
\frac{1}{\not\! p - m^0_2 + \Sigma_{22}(\not\! p)}
\Sigma_{21}(\not\! p) \Big]^{-1}, \\
\label{S22}
S_{22}(\not\! p) &=& \Big[\not\! p\, -\, m^0_2\, +\,
\Sigma_{22}(\not\! p)\, -\, \Sigma_{21}(\not\!
p) \frac{1}{\not\! p - m^0_1 + \Sigma_{11}(\not\! p)}
\Sigma_{12}(\not\! p) \Big]^{-1}, \\
\label{S12}
S_{12}(\not\! p) &=& -\, S_{11}(\not\! p)\,
\Sigma_{12}(\not\! p)\, \Big[ \not\! p\, -\, m^0_2\, +\,
\Sigma_{22}(\not\! p) \Big]^{-1} \nonumber\\ &=& -\, \Big[
\not\! p\, -\, m^0_1\, +\, \Sigma_{11}(\not\! p) \Big]^{-1}
\Sigma_{12}(\not\! p)\, S_{22}(\not\! p)\ ,\\
\label{S21}
S_{21}(\not\! p) &=& -\, S_{22}(\not\! p)\,
\Sigma_{21}(\not\! p)\, \Big[ \not\! p\, -\, m^0_1\, +\,
\Sigma_{11}(\not\! p) \Big]^{-1}\, \nonumber\\ &=& -\, \Big[
\not\! p\, -\, m^0_2\, +\, \Sigma_{22}(\not\! p)
\Big]^{-1} \Sigma_{21}(\not\! p)\, S_{11}(\not\! p)\ .
\end{eqnarray}
${}$From Eqs.\ (\ref{S12}) and (\ref{S21}), it is now easy to see that
the resummed propagator matrix is endowed with a factorization
property analogous to Eq.\ (\ref{Drel}). There is also an analogous
connection between the renormalized and unrenormalized resummed 
propagators, which may be cast into the form
\begin{equation}
\label{S_Shat}
S_{ij}(\not\! p)\ =\ (Z^{1/2}_{Lim}\, P_L\ +\ Z^{1/2}_{Rim}\, P_R )\,
\widehat{S}_{mn}(\not\! p)\, (Z^{1/2\dagger}_{Lnj}\, P_R\ +\ 
Z^{1/2\dagger}_{Rnj}\, P_L )\, ,
\end{equation}
where the caret on $S_{ij}(\not\! p)$ refers to the fact that the resummed
fermionic propagators have been OS renormalized.  By analogy, the
renormalized propagators $\widehat{S}_{ij}(\not\! p)$ may be recovered
from $S_{ij}(\not\! p)$ in Eqs.\ (\ref{S11})--(\ref{S21}), if one makes
the obvious replacements: $m^0_i\to m_i$ and $\Sigma_{ij}(\not\! p)\to
\widehat{\Sigma}_{ij}(\not\! p)$.

Employing the LSZ reduction formalism, one can derive the resummed
decay amplitude, $\widehat{S}_{i}$, of the unstable fermion $f_i\to
X$, in a way similar to what has been done for the scalar case.  More
explicitly, we have
\begin{eqnarray}
\label{LSZ3}
\widehat{S}_{i}\, u_i(p) &=& 
{ \cal T}^{amp}_{k\dots}\, (Z^{1/2}_{Lkm}\, P_L\, +\, Z^{1/2}_{Rkm}\, P_R )\,
\widehat{S}_{mn}(\not\! p)\, (Z^{1/2\dagger}_{Lnj}\, P_R\, +\, 
Z^{1/2\dagger}_{Rnj}\, P_L )\nonumber\\
&&\times (Z^{-1/2\dagger}_{Lji}\, P_R\, +\, 
Z^{-1/2\dagger}_{Rji}\, P_L )\, \widehat{S}^{-1}_{ii}(\not\! p)\, 
u_i(p)\nonumber\\
&=& S_i\,  u_i(p)\ -\ (1-\delta_{ij}) S_j\, 
\widehat{\Sigma}_{ji}(\not\! p)\, \Big[ \not\! p\, -\, m_j\, +\,
\widehat{\Sigma}_{jj}(\not\! p) \Big]^{-1} u_i(p)\, .
\end{eqnarray}
Again, $S_i$ represent the respective renormalized transition
amplitudes evaluated in the stable-particle approximation. The
amplitudes $S_i$ also include all high $n$-point functions, such as
vertex corrections. On the basis of the formalism presented here, we
shall calculate the CP asymmetries in the decays of heavy Majorana
neutrinos in Section 4.

Finally, we wish to offer a comment on other approaches, which are
used to analyze the phenomenon of CP violation through particle mixing
\cite{LS2,Paschos}. Recently, this issue has been studied in
\cite{AP}, within a S-matrix amplitude formalism related to the one
discussed in this section.  Obviously, the approach based on the
diagonalization of the effective Hamiltonian \cite{LS2,Paschos} is
very helpful to describe $\varepsilon$-type CP violation, if the
effective Hamiltonian is diagonalizable through a similarity
transformation, as is the case for the known $K^0\bar{K}^0$ system.
However, if the effective Hamiltonian has mathematically the Jordan
form, when expressed in a $K^0\bar{K}^0$-like basis, then it can be
shown to be non-diagonalizable via a similarity transformation. In this
case, the complex mass eigenvalues of the two mixed {\em non-free}
particles are exactly equal and, most importantly, CP violation
through particle mixing reaches its maximum attainable value
\cite{AP}.

To give a specific example, let us consider the following effective
Hamiltonian for the mixing-system of two nearly degenerate heavy
neutrinos $N_1$ and $N_2$:
\begin{equation}
\label{effHam}
{\cal H}(\not\! p)\ =\ \left[ \begin{array}{cc}
m_1 - \widehat{\Sigma}_{11}(\not\! p) & -\widehat{\Sigma}_{12}(\not\! p)\\
-\widehat{\Sigma}_{21}(\not\! p) & m_2 - \widehat{\Sigma}_{22}(\not\! p)
\end{array} \right]\ \approx\
\left[ \begin{array}{cc}
m_N + a - i|b|  & -ib\\
-ib^* & m_N - a - i|b| \end{array} \right],
\end{equation}
in the approximation $\not\!\! p \to m_N \approx m_1\approx m_2$.  In
Eq.\ (\ref{effHam}), the parameters $a$ and $b$ are real and complex,
respectively, and $m_1=m_N+a$, $m_2=m_N-a$. The complex parameter $b$
represents the absorptive part of the one-loop neutrino transitions
$N_i\to N_j$. Unitarity requires that the determinant of the
absorptive part of ${\cal H}(\not\!\! p)$ be non-negative. For the
effective Hamiltonian (\ref{effHam}), the corresponding determinant is
zero. Such an absorptive effective Hamiltonian naturally arises in the
one-lepton doublet model with two right-handed neutrinos. In the limit
$a\to |b|$, the two complex mass eigenvalues of ${\cal H} (\not\! p)$
are exactly degenerate and equal to $m_N-i|b|$. As has been shown in
\cite{AP}, the effective Hamiltonian cannot be diagonalized by a
non-unitary similarity transformation in this limit, {\em i.e.}, the
respective diagonalization matrices become singular.

Since our effective field-theoretic method does not involve
diagonalization of the effective Hamiltonian through a similarity
transformation, such singular situations are completely avoided,
leading to well-defined analytic expressions for the decay amplitudes.
Furthermore, whenever referring to heavy Majorana neutrino masses in
the following, we shall always imply the OS renormalized masses within
the conventional perturbation field theory, which differ from the real
parts of the complex mass eigenvalues of the effective Hamiltonian.
In the presence of a large particle mixing, the corresponding
eigenstates of the latter are generally non-unitary among themselves,
whereas the eigenvectors of the former form a well-defined unitary
basis, upon which perturbation theory is based.  In this context, it
is important to note that these field-theoretic OS renormalized masses
are those that enter the condition of CP invariance given in Eq.\ 
(\ref{CPinv}). Therefore, we can conclude that, if the two complex
mass eigenvalues of the effective Hamiltonian for the mixed heavy
Majorana neutrinos are equal, this does not necessarily entail an
equality between their respective OS renormalized masses, and, hence,
absence of CP violation as well \cite{AP}.

\setcounter{equation}{0}
\section{CP asymmetries}

It is now interesting to see how the resummation formalism presented
in Section 3 is applied to describe $\varepsilon$-type CP violation in
heavy Majorana neutrino decays. The same formalism can be used for the
inverse decays, which occur in the formulation of the Boltzmann
equations (see also Section 6). For completeness, we shall include
$\varepsilon'$-type CP violation in our analysis, which originates
entirely from the one-loop $\Phi lN$ irreducible vertex, and display
plots with numerical comparisons between the two kinds of CP-violating
contributions mentioned above. Moreover, we wish to address briefly
the issue pertaining to the problem of flavour-basis invariance of the
CP asymmetries.

%******************************************************************
%%%Figure 3
%******************************************************************
\begin{center}
\begin{picture}(300,100)(0,0)
\SetWidth{0.8}

\Vertex(50,50){2}
\Line(50,50)(90,50)\Text(70,62)[]{$N_i$}
\Line(130,50)(170,50)\Text(150,62)[]{$N_j$}
\GCirc(110,50){20}{0.9}\Text(110,50)[]{{\boldmath $\varepsilon$}}
\GCirc(180,50){10}{0.9}\Text(180,50)[]{{\boldmath $\varepsilon'$}}
\DashArrowLine(187,55)(220,80){5}\Text(225,80)[l]{$\Phi^\dagger$}
\ArrowLine(187,45)(220,20)\Text(225,20)[l]{$L$}

\end{picture}\\[0.7cm]
{\small {\bf Fig.\ 3:} ~$\varepsilon$- and  $\varepsilon'$-type CP violation
in the decays of heavy Majorana neutrinos.}
\end{center}

Let us consider the decay $N_1\to l^-\chi^+$ in a model with two-right
handed neutrinos, shown in Fig.\ 3. The inclusion of all other decay
modes will then be straightforward. To make our resummation formalism
more explicit, we shall first write down the transition amplitude
responsible for $\varepsilon$-type CP violation, denoted as ${\cal
T}^{(\varepsilon)}_N$, and then take CP-violating vertex corrections
into account. Applying Eq.\ (\ref{LSZ3}) to the heavy neutrino decays,
we have
\begin{equation}
\label{TN1eps}
{\cal T}^{(\varepsilon)}_{N_1}\ =\ h_{l1}\, \bar{u}_lP_R u_{N_1}\ -\
ih_{l2}\, \bar{u}_l P_R \Big[\not\! p - m_{N_2} + i\Sigma_{22}^{abs}
(\not\! p)\Big]^{-1} \Sigma_{21}^{abs}(\not\! p) u_{N_1}\, .
\end{equation}
In Eq.\ (\ref{TN1eps}), the absorptive part of the one-loop
transitions $N_j\to N_i$, with $i,j=1,2$, has the general form
\begin{equation}
\label{Sigabs}
\Sigma^{abs}_{ij} (\not\! p)\ =\ A_{ij}(p^2)\not\! p P_L\, +\, 
A^*_{ij}(p^2)\not\! p P_R\, ,
\end{equation}
where 
\begin{equation}
\label{Aij}
A_{ij}(p^2)\ =\ \frac{h_{l'i}h^*_{l'j}}{32\pi}\, \Big[\, \frac{3}{2}\, +\,
\frac{1}{2}\, \Big( 1-\frac{M^2_H}{p^2} \Big)^2\, \Big]\, .
\end{equation}
In the limit $M_H\to 0$, one finds the known result \cite{LS1,Paschos}
$A_{ij}=h_{l'i}h^*_{l'j}/(16\pi)$. On the other hand, the CP-transform
resummed amplitude describing the decay $N_1\to l^+\chi^-$,
$\overline{{\cal T}}^{(\varepsilon)}_{N_1}$, is written down
\begin{eqnarray}
\label{TCPN1eps}
\overline{{\cal T}}^{(\varepsilon)}_{N_1} &=& 
h^*_{l1}\, \bar{v}_{N_1}P_L v_l\ -\
ih_{l2}\, \bar{v}_{N_1} \Sigma_{12}^{abs}(-\not\! p) \Big[\, -\not\! p - 
m_{N_2} + i\Sigma_{22}^{abs} (-\not\! p)\Big]^{-1} P_L v_l\nonumber\\
&=& h^*_{l1}\, \bar{u}_lP_L u_{N_1}\ -\
ih^*_{l2}\, \bar{u}_l P_L  \Big[\not\! p - m_{N_2} + 
i\overline{\Sigma}_{22}^{abs}
(\not\! p)\Big]^{-1} \overline{\Sigma}_{21}^{abs}(\not\! p) u_{N_1}\, , 
\end{eqnarray}
where the charge-conjugate absorptive self-energy is given by
\begin{equation}
\label{SigCabs}
\overline{\Sigma}^{abs}_{ij} (\not\! p)\ =\ A_{ij}(p^2)\not\! p P_R\, +\, 
A^*_{ij}(p^2)\not\! p P_L\, .
\end{equation}
In deriving the last step of Eq.\ (\ref{TCPN1eps}), we have made use
of the known identities: $u(p,s)=C\bar{v}^T(p,s)$ and $C\gamma_\mu
C^{-1} = -\gamma^T_\mu $. The expressions in Eqs.\ (\ref{TN1eps}) and
(\ref{TCPN1eps}) may be simplified even further, if the Dirac equation
of motion is employed for the external spinors. Then, the two resummed
decay amplitudes, ${\cal T}^{(\varepsilon)}_{N_1}$ and
$\overline{{\cal T}}^{(\varepsilon)}_{N_1}$, take the simple form
\begin{eqnarray}
\label{TN}
{\cal T}^{(\varepsilon)}_{N_1} &=& \bar{u}_lP_R u_{N_1}\, 
\Big[\, h_{l1}\, -\, ih_{l2}\, \frac{m^2_{N_1}(1+iA_{22})A^*_{21}
+m_{N_1}m_{N_2}A_{21}}{m^2_{N_1}(1+iA_{22})^2 -m^2_{N_2}}\, \Big]\, ,\\
\label{TCPN}
\overline{{\cal T}}^{(\varepsilon)}_{N_1} &=&\bar{u}_lP_L u_{N_1}\, 
\Big[\, h^*_{l1}\, -\, ih^*_{l2}\, \frac{m^2_{N_1}(1+iA_{22})A_{21}
+m_{N_1}m_{N_2}A^*_{21}}{m^2_{N_1}(1+iA_{22})^2 -m^2_{N_2}}\, \Big]\, .
\end{eqnarray}
The two CP-conjugate matrix elements differ from one another in having
complex conjugate Yukawa couplings to each other and scalar currents
with opposite chirality. Furthermore, the respective transition
amplitudes involving the decays $N_2\to l^-\chi^+$, ${\cal
  T}^{(\varepsilon)}_{N_2}$, and $N_2\to l^+\chi^-$, $\overline{{\cal
    T}}^{(\varepsilon)}_{N_2}$, may be obtained from Eqs.\ (\ref{TN})
and (\ref{TCPN}), just by interchanging the indices `1' and `2'
everywhere in the above two formulas.

We shall now focus our attention on studying the $\varepsilon$- and
$\varepsilon'$-type mechanisms of CP violation in heavy Majorana
neutrino decays.  For this purpose, we define the following
CP-violating parameters:
\begin{eqnarray}
\label{epsNi}
\varepsilon_{N_i} & =& \frac{|{\cal T}^{(\varepsilon)}_{N_i}|^2\, -\,
|\overline{{\cal T}}^{(\varepsilon)}_{N_i}|^2}{
|{\cal T}^{(\varepsilon)}_{N_i}|^2\, +\, 
|\overline{{\cal T}}^{(\varepsilon)}_{N_i}|^2}\ ,\qquad \mbox{for}\ i=1,2\, ,\\
\label{epsN}
\varepsilon_N & =& \frac{|{\cal T}^{(\varepsilon)}_{N_1}|^2\, +\,
|{\cal T}^{(\varepsilon)}_{N_2}|^2\,
-\, |\overline{{\cal T}}^{(\varepsilon)}_{N_1}|^2 
\, -\, |\overline{{\cal T}}^{(\varepsilon)}_{N_2}|^2}{
|{\cal T}^{(\varepsilon)}_{N_1}|^2\, +\, |{\cal T}^{(\varepsilon)}_{N_2}|^2
\, +\, |\overline{{\cal T}}^{(\varepsilon)}_{N_1}|^2\, +\,
|\overline{{\cal T}}^{(\varepsilon)}_{N_2}|^2}\ .
\end{eqnarray}
Correspondingly, one can define the CP-violating parameters
$\varepsilon'_{N_i}$ and $\varepsilon'_{N}$, which may quantify CP
violation coming exclusively from the one-loop irreducible vertices.
In Eqs.\ (\ref{epsNi}) and (\ref{epsN}), the parameters
$\varepsilon_{N_i}$ and $\varepsilon_N$ share the common property that
do not depend on the final state that $N_i$ decays, despite the fact
that the individual squared matrix elements do.  In general, both
$\varepsilon$- and $\varepsilon'$-type contributions may not be
directly distinguishable in the decay widths $\Gamma (N_i\to
l^\pm\chi^\pm)$, unless one has $\varepsilon_{N_i}\gg
\varepsilon'_{N_i}$ and vice versa, for some range of the kinematic
parameters.  Evidently, the physical CP-violating observables we are
mainly interested in are
\begin{eqnarray}
\label{deltaNi}
\delta_{N_i} &=& \frac{\Gamma (N_i\to L\Phi^\dagger )\, -\, 
\Gamma (N_i\to L^C \Phi)}{\Gamma (N_i\to L\Phi^\dagger )\, +\, 
\Gamma (N_i\to L^C \Phi)}\ , \qquad \mbox{for}\ i=1,2\, ,\\
\label{deltaN}
\delta_N &=& \frac{\sum_{i=1}^2\Gamma (N_i\to L\Phi^\dagger )\, -\, 
\sum_{i=1}^2\Gamma (N_i\to L^C \Phi)}{\sum_{i=1}^2
\Gamma (N_i\to L\Phi^\dagger )\, +\, \sum_{i=1}^2\Gamma (N_i\to L^C \Phi)}\ , 
\end{eqnarray}
where $L$ refers to all fermionic degrees of freedom of the leptonic
isodoublet that heavy Majorana neutrinos can decay. Nevertheless, the
parameters $\varepsilon_{N_i}$, $\varepsilon_N$ and $\varepsilon'_{N_i}$ 
defined above are helpful to better appreciate the significance of the
two different mechanisms of CP violation.

To elucidate our resummation formalism further, it may be useful to
calculate the analytic form of the parameters $\varepsilon_{N_i}$ for
the interesting case of nearly degenerate heavy Majorana neutrinos.
Therefore, we shall consider the approximations $\Delta m^2_N =
m^2_{N_1} - m^2_{N_2} \ll m^2_{N_1} \sim m^2_{N_2}$ and define the
parameter $r_N = \Delta m^2_N/(m_{N_1}m_{N_2})$. Moreover, we neglect
high order Yukawa couplings of ${\cal O}( h^4_{lj} )$ at the amplitude
level.  It is then not difficult to find the approximate expressions
\begin{eqnarray}
\label{epsN1}
\varepsilon_{N_1} &\approx& \frac{\Im m ( h^*_{l1}h_{l2}h^*_{l'1}h_{l'2})}{
8\pi |h_{l1}|^2}\ \frac{r_N}{r^2_N\, +\, 4A^2_{22}}\, ,\\
\label{epsN2}
\varepsilon_{N_2} &\approx& \frac{\Im m ( h^*_{l1}h_{l2}h^*_{l'1}h_{l'2})}{
8\pi |h_{l2}|^2}\ \frac{r_N}{r^2_N\, +\, 4A^2_{11}}\, .
\end{eqnarray}
The difference between the above expressions and those obtained within
the framework of the ordinary perturbation theory is that the
regulating terms $A^2_{11}$ and $A^2_{22}$ are absent in the latter.
Clearly, if these finite width terms were not consistently taken into
account, this would cause a singular behaviour when the degeneracy
between the two heavy Majorana neutrinos is exact \cite{BR,LS1,Paschos}. 
On physical grounds, however, this should not be very surprising, since the
only natural parameter that can regulate such a singularity is the
finite width of the heavy Majorana neutrinos. Therefore, one of the
main advantages of our approach is that the dynamics of CP violation 
through heavy-neutrino mixing can be properly described by giving rise
to physically well-behaved analytic expressions. 

Another important point, also reported in \cite{Paschos1}, is the fact
that both parameters $\varepsilon_{N_1}$ and $\varepsilon_{N_2}$
contribute constructively to CP violation. Technically speaking, this can be
seen as follows. As has been mentioned above, the expression
$\varepsilon_{N_2}$ in Eq.\ (\ref{epsN2}) may be obtained from Eq.\ 
(\ref{epsN1}), if one replaces the heavy-neutrino index `1' with `2'
everywhere in that formula.  As a result, the CP-violating combination
of Yukawa couplings, $\Im m ( h^*_{l1}h_{l2}h^*_{l'1}h_{l'2})$, flips
sign, which gets compensated by a similar sign flip in the parameter
$r_N$.  Another significant feature of our analytic results in Eqs.\ 
(\ref{epsN1}) and (\ref{epsN2}) is that both $\varepsilon_{N_1}$ and
$\varepsilon_{N_2}$ vanish in the limit $m_{N_1}\to m_{N_2}$, which is
consistent with the requirement of CP invariance given in Eq.\ 
(\ref{CPinv}). We have checked that this property of CP invariance
persists even if one calculates $\varepsilon_{N_1}$ and
$\varepsilon_{N_2}$ exactly from the expressions (\ref{TN}) and
(\ref{TCPN}). In addition, we have verified that $\varepsilon_{N_2}$
($\varepsilon_{N_1}$) vanishes in the limit $m_{N_1}$ ($m_{N_2}$) $\to
0$, as is also prescribed by Eq.\ (\ref{CPinv}).  As a consequence,
our analytic expressions for $\varepsilon$-type CP violation are
indeed proportional to the flavour-basis invariant combination of CP
non-invariance ({\em cf.}\ Eq.\ (\ref{CPinv})), $m_{N_1} m_{N_2}
(m^2_{N_1}-m^2_{N_2}) \Im m ( h^*_{l1}h_{l2}h^*_{l'1}h_{l'2})$.

In order to make our analysis complete, it is important to include the
contributions from $\varepsilon'$-type CP violation, since they may be
significant for very large differences of heavy neutrino masses, {\em
  e.g.}, for $m_{N_1}-m_{N_2}\sim m_{N_1}$ or $m_{N_2}$.  In this
regime, the $\varepsilon$-type terms are suppressed by the large
virtuality of the heavy neutrino propagator and so become comparable
to the $\varepsilon'$-type terms \cite{LS1}.  It is now useful to
define the function
\begin{equation}
\label{Fxa}
F(x,\alpha)\ =\ \sqrt{x}\, \Big[\, 1\, -\, \alpha\, -\, (1+x)
\ln\Big(\frac{1-\alpha+x}{x}\Big)\, \Big]\, .
\end{equation}
If one sets $\alpha = 0$ in Eq.\ (\ref{Fxa}), then $F(x,\alpha)$
reduces to the known function $f(x)=\sqrt{x}[1-(1+x)\ln(1+1/x)]$,
found in \cite{FY}. Here, we are only interested in the $L$-violating
absorptive parts of the one-loop vertices $\chi^+lN_i$, 
$\chi^0\nu_lN_i$ and $H\nu_l N_i$, shown in Figs.\ 1(a)--(c). 
The complete analytic expressions are calculated in Appendix A.
For later convenience, we also assume that the external decaying
heavy Majorana neutrinos are off-shell. Having this in mind, 
we find the off-shell absorptive couplings 
\begin{eqnarray}
\label{eps'lN}
{\cal V}^{abs}_{\chi^+lN_i}(\not\! p) &=& -\, \frac{h^*_{l'i}h_{l'j}h_{lj}}{
16\pi\sqrt{p^2}}\ \not\! p P_L\, F\Big(\frac{m^2_{Nj}}{p^2}\ ,0\Big)\, ,\\
\label{eps'nuN}
{\cal V}^{abs}_{\chi^0\nu_lN_i}(\not\! p) &=& {\cal V}^{abs}_{H\nu_lN_i}
(\not\! p)\nonumber\\
&=& -\, \frac{h^*_{l'i}h_{l'j}h_{lj}}{
32\pi\sqrt{p^2}}\ \not\! p P_L\, \Big[\, F\Big(\frac{m^2_{Nj}}{p^2}\ ,0\Big)
\ +\  F\Big(\frac{m^2_{Nj}}{p^2}\ , \frac{M^2_H}{p^2}\Big)\, \Big]\, .
\end{eqnarray}
To make contact with the $\varepsilon'_{N_i}$ expressions existing 
in the literature \cite{FY,MAL,CEV,epsilonprime}, we compute the
$\varepsilon'$-type CP asymmetry in the conventional perturbation
theory, using Eqs.\ (\ref{eps'lN}) and (\ref{eps'nuN}) and neglecting
wave-function contributions.  Considering all decay channels for the
decaying heavy Majorana neutrino, {\em e.g.}, $N_1$, we obtain
\begin{eqnarray}
\label{eps'N1}
\varepsilon'_{N_1} &=& \frac{\Im m(h^*_{l1}h_{l2}h^*_{l'1}h_{l'2})}{
16\pi |h_{l1}|^2\, [\frac{3}{4} +\frac{1}{4}(1-M^2_H/m^2_{N_1})^2]}\
\Big\{\, \frac{5}{4}\, F\Big(\frac{m^2_{N_2}}{m^2_{N_1}}\ ,0\Big)\ +\ 
\frac{1}{4}\, F\Big(\frac{m^2_{N_2}}{m^2_{N_1}}\ , 
\frac{M^2_H}{m^2_{N_1}}\Big)\nonumber\\
&&+\, \frac{1}{4}\, \Big( 1\, -\, \frac{M^2_H}{m^2_{N_1}}\Big)^2\,
\Big[\, F\Big(\frac{m^2_{N_2}}{m^2_{N_1}}\ ,0\Big)\ +\ 
F\Big(\frac{m^2_{N_2}}{m^2_{N_1}}\ , \frac{M^2_H}{m^2_{N_1}}\Big)
\, \Big]\, \Big\}\, .
\end{eqnarray}
In the vanishing limit of the Higgs-boson mass, the above formula 
simplifies to \cite{FY,MAL,CEV,epsilonprime}
\begin{equation}
\label{eps'}
\varepsilon'_{N_1}\ =\ \frac{\Im m(h^*_{l1}h_{l2}h^*_{l'1}h_{l'2})}{
8\pi |h_{l1}|^2 }\ f\Big(\frac{m^2_{N_2}}{m^2_{N_1}}\Big)\, .
\end{equation}
Unlike $\varepsilon_{N_1}$, $\varepsilon'_{N_1}$ does not vanish in
the degenerate limit of the two heavy Majorana neutrinos $N_1$ and
$N_2$.  However, when the value of $m_{N_1}$ approaches to that of
$m_{N_2}$, the $\varepsilon'$-type part of the transition amplitude
squared for the $N_1$ decay becomes equal but opposite in sign with
the respective one of the $N_2$ decay. Thus, these two
$\varepsilon'$-type terms cancel one another, leading to a vanishing result
for the CP-violating parameter $\varepsilon'_N$ ($\not=\varepsilon'_{N_1} 
+ \varepsilon'_{N_2}$), which may be defined analogously to 
Eq.\ (\ref{epsN}).

We are now in a position to implement the vertex corrections of the Yukawa
couplings to the resummed amplitudes, ${\cal T}^{(\varepsilon)}_{N_1}$ and
$\overline{{\cal T}}^{(\varepsilon)}_{N_1}$, given in Eqs.\ (\ref{TN1eps})
and (\ref{TCPN1eps}), respectively. Taking Eqs.\ (\ref{eps'lN}) and
(\ref{eps'nuN}) into account, we find
\begin{eqnarray}
\label{TN1}
{\cal T}_{N_1} \hspace{-2pt}&=&\hspace{-2pt} \bar{u}_l P_R\, \Big\{ 
h_{l1}+i{\cal V}^{abs}_{l1} (\not\! p) -
i\Big[h_{l2}+i{\cal V}^{abs}_{l2} (\not\! p) \Big]  
\Big[\not\! p - m_{N_2} + i\Sigma_{22}^{abs}(\not\! p)\Big]^{-1} 
\Sigma_{21}^{abs}(\not\! p)\Big\} u_{N_1} ,\nonumber\\
&& \\
\label{TCPN1}
\overline{{\cal T}}_{N_1} \hspace{-2pt}&=&\hspace{-2pt} 
\bar{u}_l P_L\, \Big\{ 
h^*_{l1}+i\overline{{\cal V}}^{abs}_{l1} (\not\! p) 
  - i\Big[h^*_{l2}+i\overline{{\cal V}}^{abs}_{l2} (\not\! p) \Big] 
\Big[\not\! p - m_{N_2} + 
i\overline{\Sigma}_{22}^{abs}(\not\! p)\Big]^{-1} 
\overline{\Sigma}_{21}^{abs}(\not\! p)\Big\} u_{N_1} ,\nonumber\\
&& 
\end{eqnarray}
where we have simplified the notation of the off-shell one-loop vertices
as ${\cal V}^{abs}_{li}(\not\!\! p)$. The vertex functions
$\overline{{\cal V}}^{abs}_{li}(\not\! p)$ are the charge conjugates of
${\cal V}^{abs}_{li}(\not\! p)$ and may hence be recovered from Eqs.\
(\ref{eps'lN}) and (\ref{eps'nuN}), by taking the complex conjugate for
the Yukawa couplings and replacing $P_R$ with $P_L$.  It may not be very
convenient to present analytic expressions for the CP-violating
observables $\delta_{N_i}$, defined in Eq.\ (\ref{deltaNi}), in a rather
compact form, although their derivation from Eqs.\ (\ref{TN1}) and
(\ref{TCPN1}) is quite straightforward.  Instead, we shall compare the
numerical results obtained from our resummation approach with those found
with different methods.

Before we proceed with our numerical analysis, it is crucial to take
the out-of equilibrium constraints on heavy neutrino decays into
account.  Detailed study of the latter will be performed by solving
numerically the Boltzmann equations in Section 6. To a good
approximation however, Sakharov's third condition imposes a lower
bound on the lifetime of the decaying heavy Majorana neutrino,
$1/\Gamma_{N_i}$, which may qualitatively be given by the inequality
\begin{equation}
\label{Sakh3}
\Gamma_{N_i} (T=m_{N_i})\ \stackrel{\displaystyle <}{\sim}\
2\, H(T=m_{N_i})\, ,
\end{equation}
where $H(T)$ is the Hubble parameter 
\begin{equation}
\label{Hubble}
H(T)\ =\ 1.73\, g_*^{1/2}\, \frac{T^2}{M_{Planck}}\ .
\end{equation}
In Eq.\ (\ref{Hubble}), $g_*\approx 100-400$ represents the number of
active degrees of freedom in usual extensions of the SM and $M_{Planck} =
1.2\ 10^{19}$ GeV is the Planck mass scale.  Considering the total decay
width of the heavy neutrino, $\Gamma_{N_i}\ =\ |h_{li}|^2 m_{N_i}/(8\pi)$,
the out-of equilibrium constraint in Eq.\ (\ref{Sakh3}) yields
\begin{equation}
\label{hli_bound}
|h_{li}|^2\ \stackrel{\displaystyle <}{\sim}\
7.2 \times 10^{-14}\, \Big( \frac{m_{N_i}}{1\ \mbox{TeV}}\Big)\, .
\end{equation}
In order to have the baryon-to-entropy density ratio in the observed
ball park, {\em i.e.}, $Y_B \approx 10^{-10}\approx |\delta_{N_i}|/
g_*\, $, one must allow for CP asymmetries $|\delta_{N_i}|$ of order
$10^{-7} - 10^{-6}$. This is practically independent of the heavy
neutrino mass, for masses $m_{N_i} \stackrel{\displaystyle >}{\sim} 1$
TeV, provided the out-of equilibrium constraint on the Yukawa coupling
in Eq.\ (\ref{hli_bound}) is fulfilled.

\begin{figure}[hb]
   \leavevmode
 \begin{center}
   \epsfxsize=15.cm
   \epsffile[0 0 539 652]{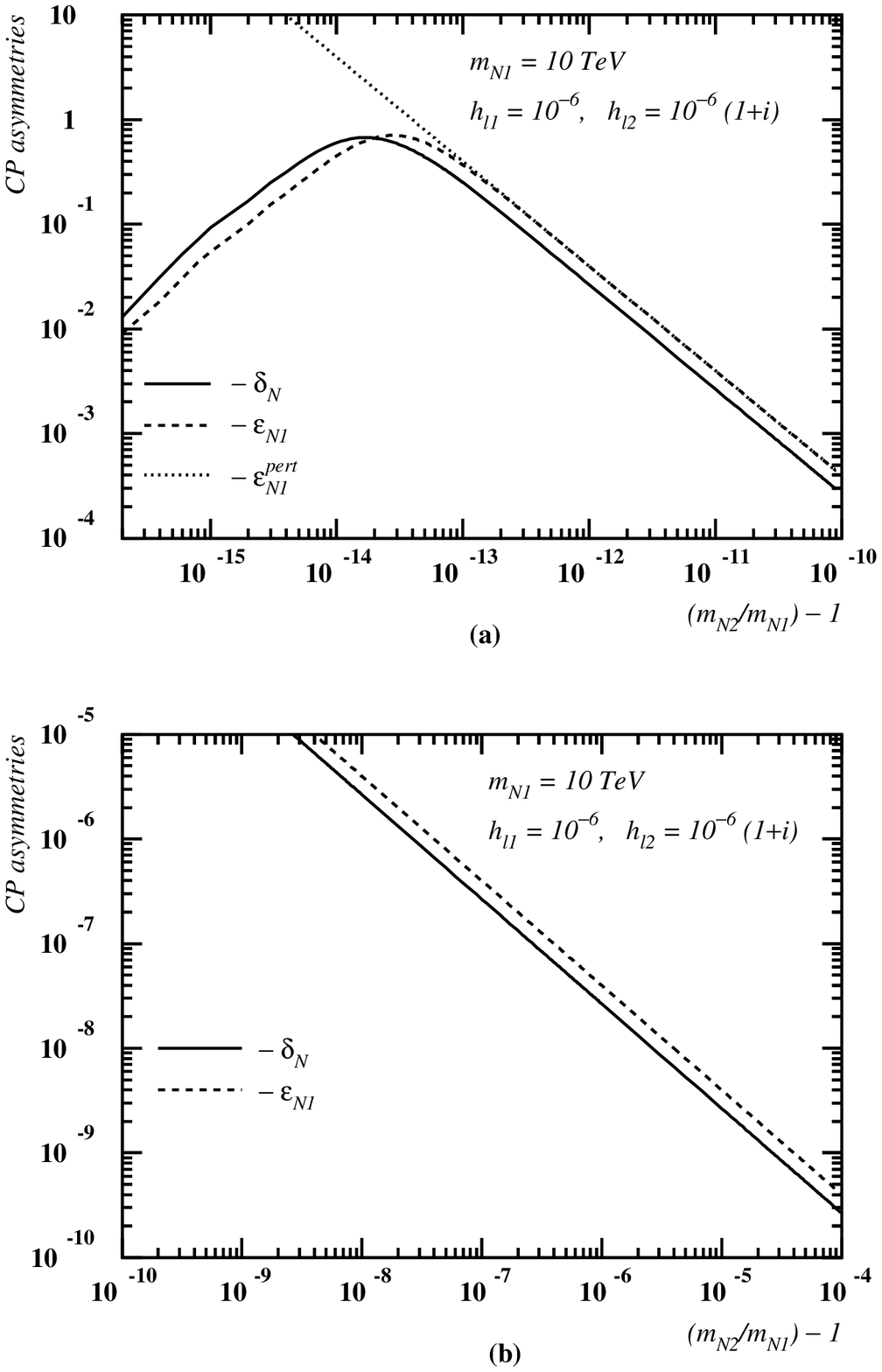}
{\small {\bf Fig.\ 4:} Numerical estimates of CP asymmetries in scenario I.}
 \end{center}
\end{figure}

\begin{figure}[ht]
   \leavevmode
 \begin{center}
   \epsfxsize=15.cm
   \epsffile[0 0 539 652]{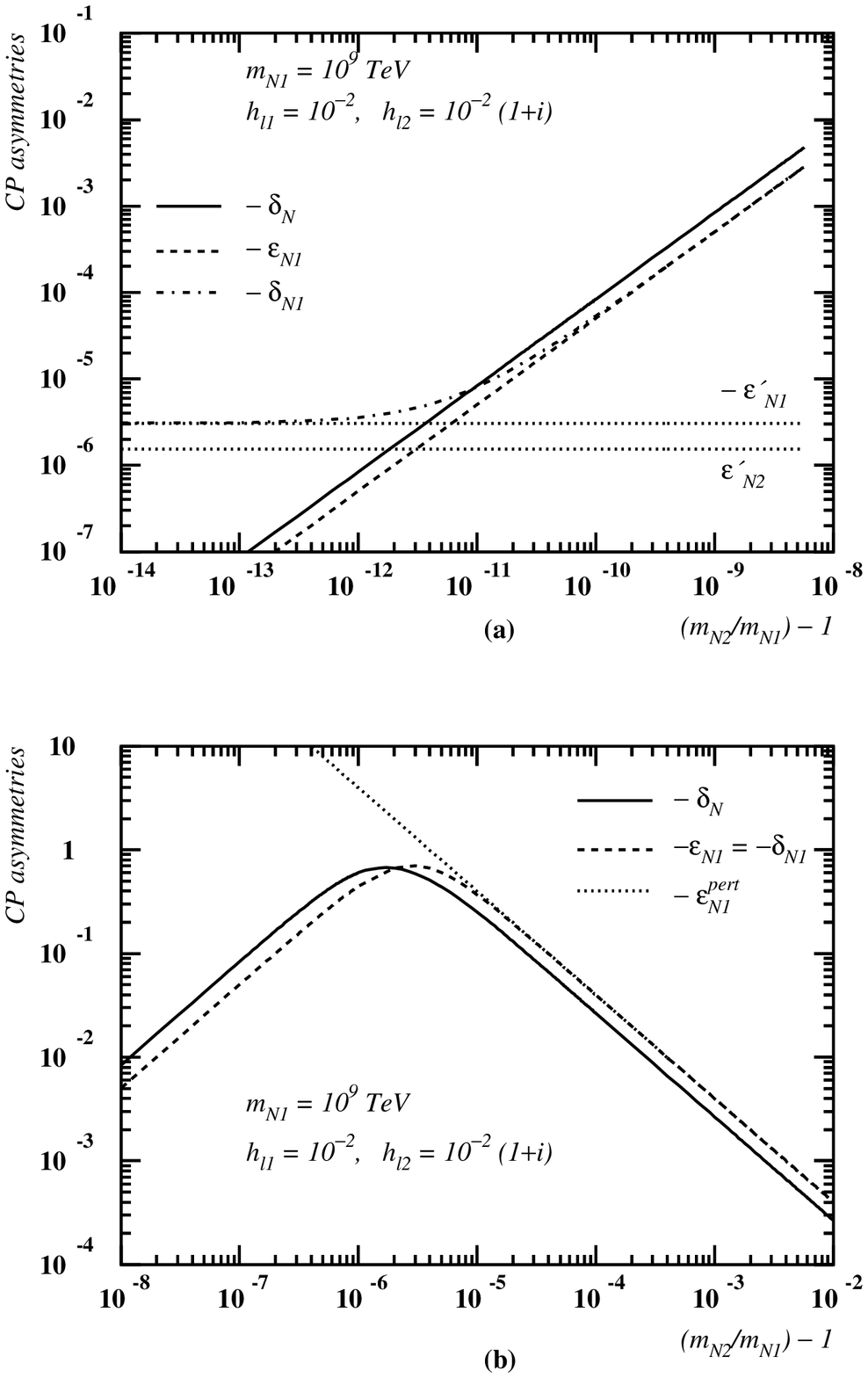}
{\small {\bf Fig.\ 5:} Numerical estimates of CP asymmetries versus
$m_{N_2}/m_{N_1} -1 $ in scenario II.}
 \end{center}
\end{figure}

\begin{figure}[hb]
   \leavevmode
 \begin{center}
   \epsfxsize=12.cm
   \epsffile[0 0 425 425]{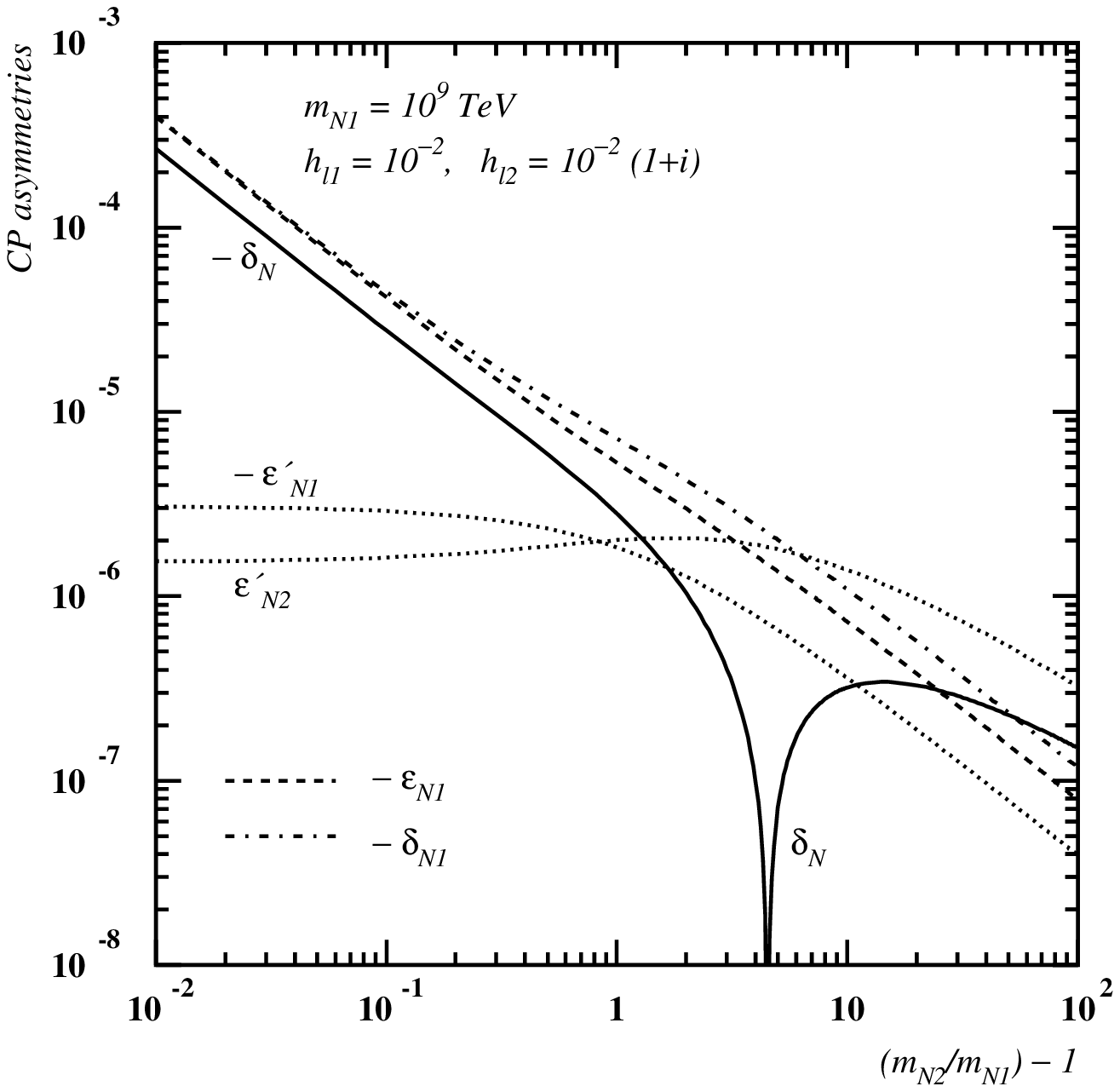}
{\small {\bf Fig.\ 6:} Numerical estimates of CP asymmetries as a function
of $m_{N_2}/m_{N_1} -1 $ in scenario II.}
 \end{center}
\end{figure}

We shall give numerical estimates of CP asymmetries in two heavy-neutrino
scenarios, which are in compliance with the out-of-equilibrium limits
derived above. For our illustrations, we analyze models, in which the
two-right handed neutrinos mix actively with one lepton family $l$ only.
Despite their simplicity, such models exhibit all the essential features
of CP violation through heavy-neutrino mixing.  Specifically, we consider
the following two scenarios:
\begin{eqnarray}
\label{scenario}
\mbox{I.} && m_{N_1}\, =\, 10\ \mbox{TeV}\, ,\qquad h_{l1}=10^{-6},\
\quad h_{l1}=10^{-6}(1+i)\, ,\nonumber\\
\mbox{II.} && m_{N_1}\, =\, 10^9\ \mbox{TeV}\, ,\qquad h_{l1}=10^{-2},\
\quad h_{l1}=10^{-2}(1+i)\, ,
\end{eqnarray}
and assume that $N_2$ is always heavier than $N_1$, {\em i.e.}, 
$m_{N_1}\leq m_{N_2}$. 

In Fig.\ 4, we display numerical estimates of the CP asymmetries
defined above as a function of the parameter $x_N = m_{N_2}/m_{N_1}
-1$ for the scenario I.  We have divided the range of values for the
parameter $x_N$ into two regions: The first region is plotted in Fig.\ 
4(a) and pertains to the kinematic domain, where resonant CP violation
due to heavy-neutrino mixing occurs. The second one, shown in Fig.\ 
4(b), represents the kinematic range, far away from the resonant
CP-violating phenomenon. The dotted line in Fig.\ 4(a) gives the
prediction of $\varepsilon_{N_1}$, obtained from Eq.\ (\ref{epsN1}) in
the conventional perturbation theory. Obviously, $\varepsilon^{pert}_{N_1}$ 
diverges for sufficiently small values of $x_N$, {\em e.g.}, $x_N <
10^{-13}$.  If resummation of the relevant fermionic self-energy
graphs is considered, the prediction for $\varepsilon_{N_1}$ is given
by the dashed lines in Fig.\ 4, which shows a maximum for $x_N\approx
10^{-13}$.  In such a case, CP violation may resonantly increase up to
order of unity \cite{APRL,AP}.  As has been mentioned above, the
physical regulating parameter of the singularity in
$\varepsilon^{pert}_{N_1}$ is the finite width of the heavy neutrino
$N_2$, which arises naturally within our field-theoretic approach.
Thus, the condition for resonant enhancement of CP violation reads:
\begin{equation}
\label{CPcond}
m_{N_1}\, -\, m_{N_2}\ \sim\ \pm\, A_{22} m_{N_2}\, =\, \frac{\Gamma_{N_2}}{2}\
\ \mbox{and/or}\quad A_{11} m_{N_1}\, =\, \frac{\Gamma_{N_1}}{2}\ . 
\end{equation}
Clearly, Fig.\ 4(a) satisfies the above condition.  However, 
the magnitude of the CP asymmetries is governed by the expression
\begin{equation}
\label{dCP}
\delta_{CP}\ =\ \frac{|\Im m ((h^*_{l1}h_{l2})^2)|}{|h_{l1}|^2 |h_{l2}|^2}\ ,
\end{equation}
which is always $\delta_{CP}\leq 1$. Evidently, both scenarios I and II
given above represent maximal cases of CP violation with $\delta_{CP}=1$.
Therefore, results for any other model may readily be read off from Figs.\
4, 5 and 6, by multiplying them with the appropriate model-dependent
factor $\delta_{CP}$.  The solid line in Fig.\ 4 gives the numerical
estimate for the CP-violating parameter $\delta_N$ in Eq.\ (\ref{deltaN}),
where $\varepsilon'$-type contributions are included.  The latter are very
small in this scenario, so as to potentially account for the BAU, {\em
e.g.}, $\varepsilon'_{N_1}\approx 10^{-16}$. Furthermore, it may be
important to stress that $\delta_N$ vanishes in the CP-invariant limit
$x_N\to 0$, as it should be on account of Eq.\ (\ref{CPinv}).

In Figs.\ 5 and 6, we give numerical estimates of the CP asymmetries in
the scenario II. The difference of this model with the scenario I is that
the $\varepsilon'$-type effects may not be negligible in the off-resonant
region, as can be seen from Figs.\ 5(a) and 6.  In particular, for values
of the parameter $x_N < 10^{-11}$ or $x_N > 1$, the individual 
$\varepsilon'_{N_1}$- and $\varepsilon'_{N_2}$-type contributions
may dominate over those of the $\varepsilon$ type.  Models with
$x_N> 1$ have extensively been discussed in the literature
\cite{FY,MAL,CEV,epsilonprime}. Numerical estimates for such models are
displayed in Fig.\ 6.  Our attention will now be focused on the domain
with $x_N<10^{-2}$.  In Fig.\ 5(a), we observe that $\varepsilon'_{N_1}$
and  $\varepsilon'_{N_2}$, represented by the dotted lines, do not vanish
in the CP-invariant limit $x_N\to 0$, as opposed to $\varepsilon_{N_1}$.
As a consequence, the CP asymmetry $\delta_{N_1}$ in Eq.\ (\ref{deltaNi}),
in which both $\varepsilon_{N_1}$- and $\varepsilon'_{N_1}$-type terms are
considered within our formalism, does not vanish either.  The reason is
that the physical CP-violating parameter in this highly degenerate mass
regime for $N_1$ and $N_2$ is the observable $\delta_N$ defined in Eq.\
(\ref{deltaN}). In fact, the quantity $\delta_N$ shares the common feature
with $\varepsilon_{N_1}$ and tends consistently to zero as $x_N\to 0$.
This fact must be considered to be one of the successes of our resummation
approach.  Again, CP violation is resonantly amplified, when the condition
in Eq.\ (\ref{CPcond}) is satisfied, as can be seen from Fig.\ 5(b).
Finally, we must remark that $-\delta_N$ has a zero point and eventually
becomes negative for $x_N\gg 1$, as is plotted in Fig.\ 6.  Nevertheless,
this result should be viewed with great caution.  The actual reason is
that the effect of the different dissipative Boltzmann factors multiplying
the decay rates of the heavy Majorana neutrinos $N_1$ and $N_2$ must be
considered in the definition of $\delta_N$ in Eq.\ (\ref{deltaN}). These
phenomena will be taken into account in Section 6, in which we shall write
down and solve numerically the relevant Boltzmann equations.

Our computation of the CP asymmetries has been carried out in the physical
basis, in which the heavy neutrino mass matrix is non-negative and
diagonal. One might, however, raise the question whether our analytic
results would have been modified if we had chosen a different basis other than
the mass basis. The best way to address this question is to discuss first
how our expressions would change under a basis transformation. Let us assume a
model with two-right handed neutrinos for simplicity and imagine that we
wish to perform our calculations in a basis, in which the two heavy
Majorana neutrinos, $n_1$ and $n_2$, say, span a non-diagonal mass matrix.
The heavy Majorana neutrinos $n_1$ and $n_2$ are then related to the
neutrinos $N_1$ and $N_2$, defined in the mass basis, through the unitary
transformation:
\begin{equation}
\label{Un}
\left(\begin{array}{c} n_1\\ n_2 \end{array}\right)_R\ =\ 
U^n\, \left(\begin{array}{c} N_1 \\ N_2 \end{array}\right)_R\, ,\qquad
\left(\begin{array}{c} n_1\\ n_2 \end{array}\right)_L\ =\ 
U^{n*}\, \left(\begin{array}{c} N_1 \\ N_2 \end{array}\right)_L\, .
\end{equation}
The $2\times 2$ unitary matrix $U^n$ relates the physical and non-diagonal
mass matrices in the following way:
\begin{equation}
\label{Mn_Mnu}
\widehat{M}^\nu \ =\ U^{nT}\, M^n\, U^n\, .
\end{equation} 
It is now helpful to see how the various kinematic parameters
transform under the action of $U^n$. In particular, the inverse
propagator matrix $S^{-1}_N(\not\! p)$ undergoes a change 
depending on $S^{-1}_n (\not\! p)$, which is given by
\begin{equation}
\label{SnSN_1}
S^{-1}_N (\not\! p) \ =\ (U^{nT}P_R\, +\, U^{n\dagger}P_L )\, 
S^{-1}_n (\not\! p)\, (U^nP_R\, +\, U^{n*}P_L )\, .
\end{equation}
This in turn implies that 
\begin{equation}
\label{SnSN}
S_N (\not\! p) \ =\ (U^{n\dagger}P_R\, +\, U^{nT}P_L )\, 
S_n (\not\! p)\, (U^{n*}P_R\, +\, U^n P_L )\, ,
\end{equation}
where $S_N (\not\! p)$ and $S_n (\not\! p)$ are the $2\times 2$ propagator
matrices, evaluated in the two different heavy neutrino bases. Finally,
the Yukawa couplings $h = (h_{l1},\ h_{l2})$, defined in the mass basis,
are related to the Yukawa couplings $h^n = (h^n_{l1},\ h^n_{l2})$ of the
non-diagonal basis, via the unitary rotation: $h =h^n U^n$.

Given the afore-mentioned transformation of the Yukawa couplings and that
of the resummed propagators in Eq.\ (\ref{SnSN}) under a basis rotation,
it is not difficult to show that the propagator part of the S-matrix
amplitude for the process, {\em e.g.}, $l^-\chi^+\to l^+\chi^-$,
\begin{displaymath}
{\cal T}(l^-\chi^+\to l^+\chi^-)\ \propto\ 
                             \mbox{Tr} [h P_R S_N(\not\! p) P_R h^T]\, , 
\end{displaymath}
is rotational invariant under $U^n$, {\em i.e.}, 
\begin{displaymath}
\mbox{Tr} [h P_R S_N(\not\! p) P_R h^T]\ =\ 
                       \mbox{Tr} [h^n P_R S_n(\not\! p) P_R h^{nT}]\, ,
\end{displaymath}
where the trace should be taken over the product of spinor and flavour
matrices as well.  In fact, this avenue has been followed in \cite{AP}. If
decays of particles are considered however, such as heavy Majorana
neutrino decays, the above flavour-basis invariance is not manifest.  The
reason is that, within the LSZ formalism, the S-matrix amplitude is
obtained by amputating the external legs of the Green functions with
inverse propagators defined in the mass basis and any basis transformation
will affect the S-matrix expression for the decaying particle. 

Therefore, it is essential to introduce a method that always makes
reference to the decay of the neutrinos in the diagonal mass basis. To
accomplish this, one has to truncate the Green function, {\em e.g.},
${\cal T}_n^{amp} S_n (\not\!  p)$, for the decay $n_i\to l^-\chi^+$,
in the following scheme:
\begin{equation}
\label{FBI2}
\widehat{S}_i u_{N_i}(p)\ =\
{\cal T}_n^{amp} S_n (\not\! p) (U^{n*}P_R + U^n P_L)
(Z^{-1/2\dagger}_LP_R + Z^{-1/2\dagger}_R P_L)_i (\widehat{S}_N)^{-1}_{ii}
(\not\! p)\, u_{N_i}(p)\, ,
\end{equation}
where $Z_L=Z^*_R$ due to the Majorana property of the heavy neutrinos
and summation over not displayed indices is implied.  Note that ${\cal
T}_n^{amp}$ in Eq.\ (\ref{FBI2}) is calculated in the non-diagonal
basis.  We must remark that our field-theoretic approach to solving
the problem of flavour-basis invariance leads to non-vanishing CP
asymmetries proportional to the flavour-basis independent combination
of CP invariance given in Eq.\ (\ref{CPinv}). In particular, in the limit 
$(m_{N_2}-m_{N_1})\to 0$, the vanishing of $\delta_N$ in Figs.\ 4(a) and
5(a)  explicitly demonstrates that our numerical estimates describe
genuine effects of CP violation.

\setcounter{equation}{0}
\section{Low-energy constraints}

If heavy Majorana neutrinos are not much heavier than few TeV
\cite{HLP}, these novel particles may then be produced at high-energy
$ee$ \cite{PRODee}, $ep$ \cite{PRODep}, and $pp$ colliders
\cite{PRODpp}, whose subsequent decays can give rise to distinct
like-sign dilepton signals. If heavy Majorana neutrinos are not
directly accessed at high-energy machines, they may have significant
non-decoupling quantum effects on lepton-flavour-violating decays of
the $Z$ boson \cite{KPS_Z,BSVMV}, the Higgs particle ($H$)
\cite{APhiggs}, and the $\tau$ and $\mu$ leptons \cite{IP}. Their
presence may cause breaking of universality in leptonic diagonal
$Z$-boson \cite{BKPS} and $\pi$ decays \cite{KP} or influence
\cite{SB&AS} the size of the electroweak oblique parameters $S$, $T$
and $U$ \cite{STU}. In fact, there exist many observables \cite{LL},
including the $\tau$-polarization asymmetries, neutrino-counting
experiments at the CERN Large Electron Positron Collider (LEP1) or at
the Stanford Linear Accelerator (SLC), to which Majorana neutrinos may
have sizeable contributions. However, if the out-of-equilibrium
constraints are imposed on all lepton-families of the model ({\em
cf.}\ Eq.\ (\ref{hli_bound})), all these non-SM effects mentioned
above are estimated to be extremely small at the tree or one-loop
level.  For example, typical tree-level terms breaking charged-current
universality in $\pi$ decays due to heavy Majorana masses depend
linearly on the ratio $|h_{li}v|^2/m^2_{N_i}$ and are therefore
very suppressed. Moreover, one-loop induced flavour-changing decays
of the $Z$ boson are also very small of order
\begin{equation}
\label{nonSM}
\frac{\alpha_w}{4\pi}\, \frac{|h_{li} v|^2}{m^2_{N_i}}\, \frac{v^2}{M^2_W}\
\stackrel{\displaystyle <}{\sim}\ 10^{-15}\, \times\, \Big( 
\frac{1\ \mbox{TeV}}{m_{N_i}}\Big)^2\, ,
\end{equation}
at the amplitude level, with $\alpha_w$ being the SU(2)$_L$
electroweak fine structure constant. 

It is known that below the critical temperature $T_c$ of the first-order
electroweak phase transition, the Higgs boson acquires a non-vanishing
VEV \cite{KRS}. This leads to non-zero light neutrino masses, which may be
obtained from the five dimensional operator
\begin{equation}
L^T\langle \Phi^T \Phi\rangle\, L\, h^T(\widehat{M}^\nu)^{-1} h\  \approx\
\bar{\nu}_{lL}^C \nu_{lL}\ \sum_{i=1}^2\,h^2_{li}\, \frac{v^2}{m_{N_i}}\, .
\end{equation}
However, for the scenarios I and II, the light-neutrino masses
generated at temperatures $T\ll T_c$ are much below the cosmological
limit, {\em i.e.}, $m_\nu \ll 10$ eV. Obviously, all the
afore-mentioned limits cannot jeopardize the viability of our minimal
new-physics scenarios.
%******************************************************************
%%%Figure 7
%******************************************************************
\begin{center}
\begin{picture}(200,150)(0,0)
  \SetWidth{0.8}

\ArrowLine(50,80)(80,80)\Text(50,90)[l]{$e^-$}
\Line(80,80)(110,80)\Text(95,80)[]{{\boldmath $\times$}}
\Text(95,68)[]{$N_i$}
\ArrowLine(140,80)(110,80)\Text(125,90)[]{$e^-$} 
\Line(140,80)(170,80)\Text(155,80)[]{{\boldmath $\times$}}
\Text(155,92)[]{$N_j$}
\ArrowLine(170,80)(200,80)\Text(200,90)[r]{$e^-$}
\DashArrowArcn(110,80)(30,180,360){5}
\DashArrowArc(140,80)(30,180,270){5}
\DashArrowArc(140,80)(30,270,360){5}
\Photon(140,50)(140,20){3}{4}\Text(145,20)[l]{$\gamma$}
\Text(110,115)[b]{$\chi^-$}
\Text(118,60)[tr]{$\chi^-$}\Text(163,60)[lt]{$\chi^-$}

\end{picture}\\[0.5cm]
{\small {\bf Fig.\ 7:} ~Typical two-loop diagram contributing to the EDM 
of electron.}
\end{center}

New-physics interactions may also give large contributions to the EDM of
electron, whose experimental upper bound is $(d_e/e) < 10^{-26}$ cm
\cite{PDG}.  In particular, this bound is crucial, since it may impose a
constraint on CP-violating operators similar to those that induce CP
violation in heavy Majorana neutrino decays. In our SM extension with
right-handed neutrinos, the contribution to the EDM of electron arises at
two loops \cite{Ng2}. A typical diagram that gives rise to an EDM for
the electron,
\begin{equation}
\label{EDM}
{\cal L}_{d}\ =\ ie\, \Big(\frac{d_e}{e}\Big)\, \bar{e}\, \sigma_{\mu\nu}
\gamma_5\, e\,  \partial^\mu A^\nu\, ,
\end{equation}
is shown in Fig.\ 7. Following the semi-quantitative prescription in
\cite{Ng2} by making naive loop counting for $m_{N_2},\ m_{N_1}\gg
M_W$, we find
\begin{eqnarray}
\label{EDMaj}
(\frac{d_e}{e}\Big) &\sim& \Im m(h_{1e}h^*_{2e}h_{1l}h^*_{2l})\,
\frac{\alpha^2_w}{16\pi^2}\, \frac{m_e}{M^2_W}\, \frac{v^4}{M^4_W}\,
\frac{m_{N_1}m_{N_2}(m^2_{N_1}-m^2_{N_2})}{(m^2_{N_1} + m^2_{N_2})^2}\, 
\ln\Big(\frac{m_{N_1}}{M_W}\Big)\nonumber\\
&\sim& 10^{-24}\, \mbox{cm}\, \times\, \Im m(h_{1e}h^*_{2e}h_{1l}h^*_{2l})\, 
\frac{m_{N_1}m_{N_2}(m^2_{N_1}-m^2_{N_2})}{(m^2_{N_1} + m^2_{N_2})^2}\, 
\ln\Big(\frac{m_{N_1}}{M_W}\Big)
\, ,
\end{eqnarray}
where the remaining factor depending on the masses of the heavy Majorana
neutrinos only in Eq.\ (\ref{EDMaj}) is usually less than one. It is then
obvious that the above EDM limit may be important, if $|h_{li}|>0.1$, {\em
i.e.}, for super-heavy Majorana neutrinos with $m_{N_i}>10^{11}$ TeV.  On
the other hand, stability of the Higgs potential under radiative corrections 
requires that $|h_{li}|={\cal O}(1)$ \cite{HLP}. Nevertheless, the EDM
bound derived above gets less restrictive when the mass difference between
$N_1$ and $N_2$ is much smaller than the sum of their masses, {\em i.e.},
$(m_{N_1}-m_{N_2})/(m_{N_1}+m_{N_2})\ll 1$, and is practically absent for
values of the parameter $x_N < 10^{-3}$. As a consequence, in our
analysis, we have only considered scenarios with $|h_{l1}|,\ |h_{l2}|
\stackrel{\displaystyle <}{\sim} 10^{-2}$, on which the EDM constraint
in Eq.\ (\ref{EDMaj}) is still weak. 

In the above discussion of low-energy effects, we have assumed that
the out-of equilibrium constraint given in Eq.\ (\ref{hli_bound})
applies to all lepton families. As we have seen in Section 4, one
lepton family is sufficient to get the CP asymmetries required for the
BAU. Furthermore, sphaleron interactions preserve the individual
quantum numbers $B/3-L_i$, {\em e.g.}, $B/3 - L_e$. Also, an excess in
$L_e$ will be converted into the observed asymmetry in $B$. Most
interestingly, the so-generated BAU will not be washed out, even if
operators that violate $L_\mu$ and $L_\tau$ are in thermal equilibrium
provided that possible $L_e-L_\mu$- and $L_e-L_\tau$-violating
interactions are absent \cite{Dreiner/Ross}. For instance, this can be
realized if there are two right-handed neutrinos that mix with the
electron family only and produce the BAU, and the remaining
isosinglet neutrinos strongly mix with the $\mu$ and $\tau$ families.
This class of heavy Majorana neutrino models may predict sizeable
new-physics phenomena, which have been mentioned above and can be
probed in laboratory experiments.

\setcounter{equation}{0}
\section{Boltzmann equations}

In this section, we write down the relevant Boltzmann equations
\cite{KW,EWK&SW,KT,HT}, which determine the time evolution of the
lepton-number asymmetries.  Then, we solve these equations numerically
and present results for the expected BAU within the two different
democratic-type scenarios I and II discussed in Section 4.  Lastly, we
give estimates of the finite temperature effects, which may have an
impact on the resonant phenomenon of CP violation due to mixing.

The Boltzmann equations describing the time evolution of the lepton
asymmetry for a system with two heavy Majorana neutrinos are given 
by \cite{KT,MAL}
\begin{eqnarray}
\label{BENi}
\frac{dn_{N_i}}{dt}\, +\, 3Hn_{N_i} &=& 
-\, \Big( \frac{n_{N_i}}{n^{eq}_{N_i}}\, -\, 1\Big)\, \gamma_{N_i}\, ,\\ 
\label{BElept}
\frac{dn_L}{dt}\, +\, 3Hn_{L} &=& \sum\limits_{i=1}^2\, \Big[\, \delta_{N_i}
\, \Big( \frac{n_{N_i}}{n^{eq}_{N_i}}\, -\, 1\Big)\, -\, \frac{n_L}{2n^{eq}_l}
\, \Big]\, \gamma_{N_i}\ -\
\frac{n_L}{n^{eq}_l}\, \gamma_{\sigma}\, ,
\end{eqnarray}
where $n_{N_i}$, $n_L=n_l-n_{\bar{l}}$ are the densities of the number
of $N_i$ and the lepton-number asymmetry, respectively, and
$n^{eq}_{N_i}$ and $n^{eq}_l$ are their values in thermal equilibrium.
The Hubble parameter $H=(dR/dt)/R$ determines the expansion rate of
the Universe and also depends on the temperature $T$, through the
relation in Eq.\ (\ref{Hubble}).  In Eqs.\ (\ref{BENi}) and
(\ref{BElept}), $\gamma_{N_i}$ and $\gamma_{\sigma}$ are the collision
terms given by
\begin{eqnarray}
\label{gNi}
\gamma_{N_i}& =& n^{eq}_{N_i}\, \frac{K_1 (m^2_{N_i}/T)}{K_2(m^2_{N_i}/T)}\,
\Gamma_{N_i}\, ,\\
\label{gsigma}
\gamma_\sigma &=& \frac{T}{8\pi^4}\, \int_0^\infty\, ds\, s^{3/2}\,
K_1 (\sqrt{s}/T)\, \sigma'(s)\, .
\end{eqnarray}
Here, $K_1(z)$ and $K_2(z)$ are the modified Bessel functions defined
in \cite{MA&IAS}. In addition, $\Gamma_{N_i}$ and $\sigma'(s)$ are
respectively the usual $T=0$ expressions for the total decay-width of
$N_i$ and the cross section of the $2\to 2$ scatterings, involving the
$L$ and $\Phi$ states, which are taken here to be massless.  The
latter comprises the scatterings, {\em i.e.}, $L^C\Phi\to
L\Phi^\dagger$ and its CP-conjugate process $L\Phi^\dagger\to
L^C\Phi$. In fact, the cross section $\sigma'(s)$ is calculated by
subtracting all those real intermediate contributions that have
already been taken into account in the direct and inverse decays of
heavy Majorana neutrinos \cite{EWK&SW}. Therefore, $\gamma_{\sigma}$
may be regarded as an additional CP-conserving depletion term, which
can be shown to be of order $h^4_{li}$ in the Yukawa couplings, {\em
  i.e.}, one formally finds that $\gamma_{\sigma}\sim \gamma^2_{N_i}$
in the narrow width approximation \cite{KT}.

In writing down Eqs.\ (\ref{BENi}) and (\ref{BElept}), several
applicable assumptions have been made, which are also reviewed in
Ref.\ \cite{KT}.  First, we have considered the
Friedmann-Robertson-Walker model in the non-relativistic limit.
Second, we have adopted the Maxwell-Boltzmann statistics, which is a
good approximation in the absence of effects that originate from Bose
condensates or arise due to degeneracy of many Fermi degrees of
freedom.  Third, we have assumed that the lepton and Higgs weak
isodoublets, $L$ and $\Phi$, are practically in thermal equilibrium,
and neglected high orders in $n_L/n^{eq}_l$ and $\delta_{N_i}$. In
this context, it has also been assumed that the different particle
species are in kinetic equilibrium, {\em i.e.}, the particles may
rapidly change their kinetic energy through elastic scatterings but
the processes responsible for a change of the number of particles are
out of equilibrium. These out-of equilibrium reactions are described
by the Boltzmann equations (\ref{BENi}) and (\ref{BElept}).  Finally,
there may exist additional contributions to the Boltzmann equations
\cite{MAL}, coming from processes, such as $N_iL\to \Phi^*\to Q_i
\bar{t}_R$, $N_i Q_i\to L \bar{t}_R$, where $Q_i$ ($i=1,2,3$) denotes
the usual quark isodoublets in the SM. These reactions as well as
those of the kind $\Phi\Phi^\dagger\to LL^C$ are still very weak to
wash out the BAU generated by the direct heavy Majorana neutrino
decays \cite{MPl}, as long as the out-of equilibrium constraint on the
Yukawa couplings in Eq.\ (\ref{hli_bound}) is imposed. Hence, we have
neglected these small depletion terms.

Before we evaluate numerically the Boltzmann equations written above, it will
prove helpful to make the following substitutions:
\begin{equation}
x\ =\ \frac{m_{N_1}}{T}\ ,\qquad t\ =\ \frac{1}{2H(T)}\ =\ 
\frac{x^2}{2H(x=1)}\ ,
\end{equation}
which is a good approximation for the radiation dominated phase of the
Universe.  We assume the heavy neutrino mass hierarchy $m_{N_1}\leq
m_{N_2}$ for the two-right handed neutrino scenarios I and II, given
in Section 4.  Furthermore, we introduce the quantities
$Y_{N_i} = n_{N_i}/s$ and $Y_L = n_L/s$, where $s$ is the entropy density.
In an isentropically expanded Universe, the entropy density has the
time dependence $s(t)=\mbox{const.}\times R^{-3}(t)$ and may be
related to the number density of photons, $n_\gamma$, as $s=g_*
n_\gamma$, where $g_*$ is given after Eq.\ (\ref{Hubble}). For our
discussion, it will be more convenient to define the parameters
\begin{equation}
\label{Kparam}
K\ =\ \frac{K_1(x)}{K_2(x)}\, \frac{\Gamma_{N_1}}{H(x=1)}\ ,
\qquad \gamma\ =\ \frac{K_2(x)K_1(\xi x)}{K_1(x)K_2(\xi x)}\, 
\frac{\Gamma_{N_2}}{\Gamma_{N_1}}\ ,
\end{equation}
with $\xi = m_{N_2}/m_{N_1}$. Making use of the above definitions and
relations among the parameters, we obtain the Boltzmann equations for
the new quantities $Y_{N_1}$, $Y_{N_2}$ and $Y_L$, {\em viz.}
\begin{eqnarray}
\label{BEYN1}
\frac{dY_{N_1}}{dx} &=& -\, (Y_{N_1} - Y^{eq}_{N_1}) Kx^2\, ,\\
\label{BEYN2}
\frac{dY_{N_2}}{dx} &=& -\, (Y_{N_2} - Y^{eq}_{N_2}) \gamma Kx^2\, ,\\
\label{BEYL}
\frac{dY_L}{dx} &=& \Big[\, (Y_{N_1}-Y^{eq}_{N_1})\delta_{N_1}\, +\, 
(Y_{N_2} - Y^{eq}_{N_2} )\gamma\delta_{N_2}\, -\, \frac 12 g_* Y_L
(Y^{eq}_{N_1}+\gamma Y^{eq}_{N_2})  \nonumber\\
&&-\, g_* Y_L Y^{eq}_{N_1} \frac{\gamma_\sigma}{\gamma_{N_1}}\, 
\Big]\, Kx^2\, .
\end{eqnarray}
In our numerical analysis, we shall neglect the Yukawa coupling
suppressed term in Eq.\ (\ref{BEYL}), which is proportional to
$\gamma_{\sigma}$, since $\gamma_{\sigma} \ll \gamma_{N_i}$.
Moreover, the heavy-neutrino number-to-entropy densities in
equilibrium $Y^{eq}_{N_i}(x)$ are given by \cite{EWK&SW}
\begin{equation}
\label{YeqN}
Y^{eq}_{N_1}(x)\ =\ \frac{3}{8g_*}\, \int_{x}^\infty\, 
dz\, z\, \sqrt{z^2-x^2}\, e^{-z}\ =\ \frac{3}{8g_*}\, x^2\, K_2(x)\, , 
\end{equation}
and $Y^{eq}_{N_2}(x)=Y^{eq}_{N_1}(\xi x)$. The differential equations
(\ref{BEYN1})--(\ref{BEYL}) are solved numerically, using the initial
conditions:
\begin{equation}
\label{InBE}
Y_{N_1}(0)\ =\ Y_{N_2}(0)\ =\  Y^{eq}_{N_1}(0)\ =\ Y^{eq}_{N_2}(0)
\quad \mbox{and}\quad Y_L(0)=0\, .
\end{equation}
These initial conditions merely reflect the fact that our Universe
starts evolving from a lepton symmetric state, in which the heavy
Majorana neutrinos are originally in thermal equilibrium. After the
evolution of the Universe until temperatures much below $m_{N_1}$, a
net lepton asymmetry has been created. This lepton asymmetry will then
be converted into the BAU via the sphalerons. During a first order
electroweak phase transition, the produced excess in $L$ will lead to
an excess in $B$, which is given by \cite{BS,HT}
\begin{equation}
\label{YB_YL}
Y_B\ =\  \frac{8N_g\, +\, 4N_H}{22N_g\, +\, 13N_H}\, Y_{B-L}\ \approx\
-\, \frac{1}{3}\, Y_L\, ,
\end{equation}
where $Y_B=n_B/s$, $N_g$ is the number of generations and $N_H$ is the
number of Higgs doublets. The observed BAU is $Y^{obs}_B = (0.6 -
1)\times 10^{-10}$ \cite{KT}, which corresponds to an excess of
leptons $-Y^{obs}_L \approx 10^{-9} - 10^{-10}$. In the latter
estimate, other alternatives for generating the BAU are also
considered, which may arise from the conversion of an individual lepton
asymmetry \cite{Dreiner/Ross} only, {\em e.g.}, $L_e$, into the BAU.

\begin{figure}[ht]
   \leavevmode
 \begin{center}
   \epsfxsize=15.cm
   \epsffile[0 0 539 652]{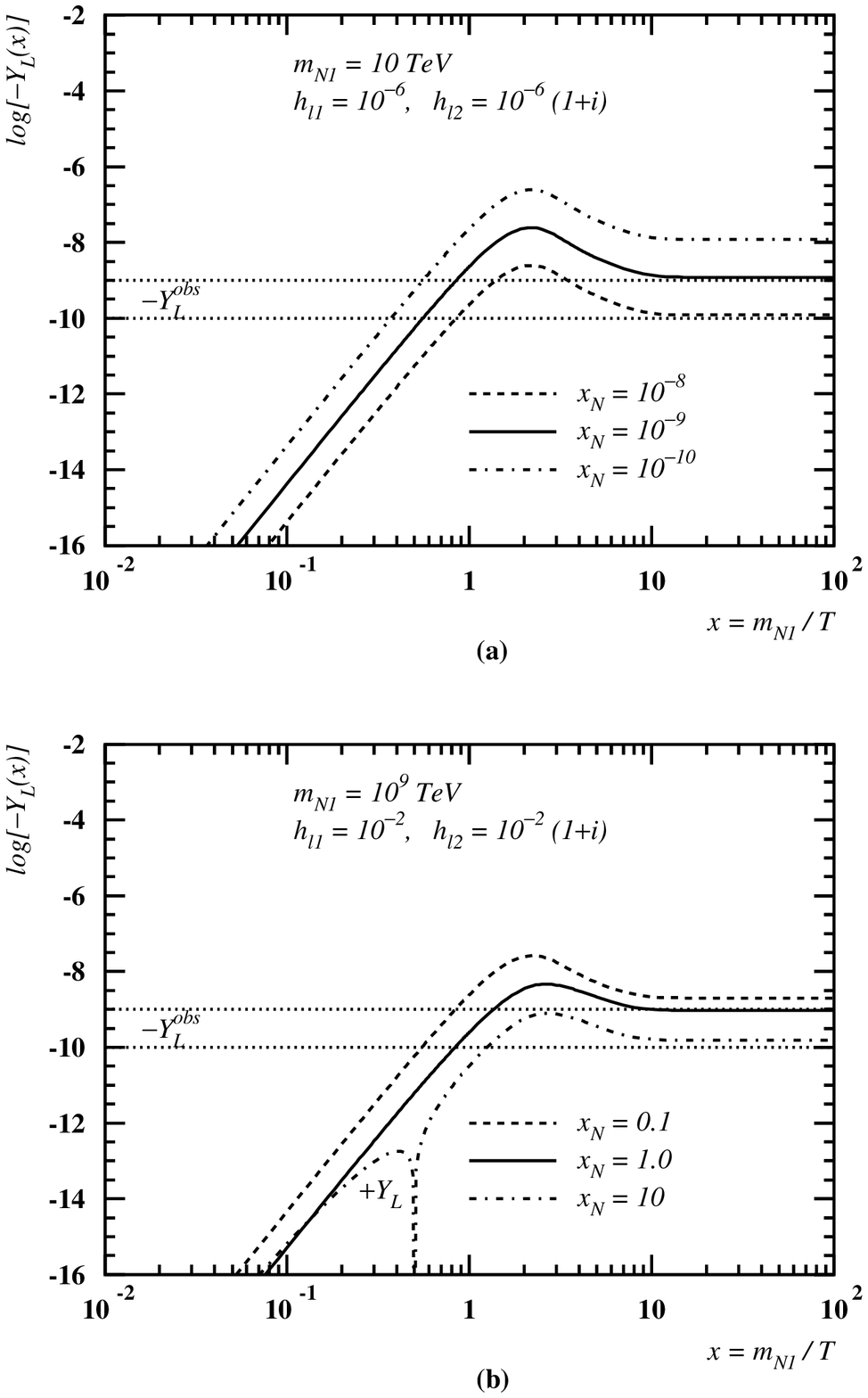}
{\small {\bf Fig.\ 8:} Lepton asymmetries for selected heavy Majorana 
neutrino scenarios.}
 \end{center}
\end{figure}

In Fig.\ 8, the observed range for $Y_L$, $Y^{obs}_L$, is indicated
with two confining horizontal dotted lines.  Furthermore, we display
our numerical estimates of $Y_L(x)$ as a function of the parameter
$x=m_{N_1}/T$, for selected heavy Majorana neutrino scenarios, stated
in Eq.\ (\ref{scenario}). Specifically, Fig.\ 8(a) shows explicitly
the dependence of $Y_L$ on $x$, for the three different values of the
parameter $x_N=m_{N_2}/m_{N_1} - 1 = 10^{-8},\ 10^{-9}$ and $10^{-10}$
in scenario I. In this scenario, the lightest heavy Majorana neutrinos
$N_1$ has a mass $m_{N_1}=10$ TeV and the values of the Yukawa
couplings are $h_{l1} = 10^{-6}$ and $h_{l2} = 10^{-6} (1+i)$. The
parameter $x_N$ is a measure of the degree of mass degeneracy for
$N_1$ and $N_2$. For comparison, it is worth mentioning that the
degree of mass degeneracy between $K_L$ and $K_S$ is of order
$10^{-15}$, which is by far smaller than the one considered here.
Since $\varepsilon'$-type CP violation is very small in scenario I, as
has already been discussed in Section 4, one has to rely on CP
violation through heavy Majorana neutrino mixing. We find that for
small heavy neutrino mass splittings determined by $x_N$, with $x_N$
being in the range between $10^{-9}$ and $10^{-8}$, a sufficiently
large lepton (baryon) asymmetry can be generated. The significance of
our $\varepsilon$-type CP-violating mechanism may be seen from the
fact that in democratic-type scenarios, {\em i.e.}, in models with all
Yukawa couplings being of the same order as those considered here,
heavy Majorana neutrinos with masses as low as 1 TeV can still be
responsible for the excess of baryons, found by observational
measurements. Of course, for larger $x_N$ values, {\em e.g.}, $x_N >
10^{-8}$, the BAU is getting much smaller than $Y^{obs}_L$.
Furthermore, we have also checked that CP violation and hence BAU
vanishes in the limit $x_N\to 0$, {\em i.e.}, when the two OS
renormalized heavy-neutrino masses $N_1$ and $N_2$ are exactly equal,
as it should be on account of the CP invariance condition in Eq.\ 
(\ref{CPinv}).

Fig.\ 8(b) gives numerical estimates of $Y_L$ as a function of $x$,
for the scenario II. In this model, we have chosen $m_{N_1} = 10^9$
TeV, and $h_{l1} = 10^{-2}$ and $h_{l2} = 10^{-2} (1 + i)$. We also
present results for three different values of the parameter $x_N$,
$x_N = 0.1,\ 1$ and 10. In this large-$m_{N_1}$ scenario, a high
degree of degeneracy for $N_1$ and $N_2$ is not required in order to
get sufficient CP violation for the BAU. In fact, the $\varepsilon$-
and $\varepsilon'$-type contributions to the decays of heavy Majorana
neutrinos are of comparable order and should both be taken into
account. Again, one finds numerically an appreciable excess of
leptons, $Y_L$, within the observed range $Y^{obs}_L$. For the
scenario with $x_N=10$, we obtain positive values for $Y_L$ up to
$x\simeq 0.5$. This small excess of leptons is rapidly erased by the
$N_1$ heavy neutrino decays, which are almost in thermal equilibrium.
At lower temperatures, {\em i.e.}, for $x \gg 0.5$, the heaviest heavy
Majorana neutrino $N_2$ gets decoupled from the system, and out-of
equilibrium $N_1$ decays will eventually produce a non-zero value for
$Y_L$ at the observable level. Since the $\varepsilon$- and
$\varepsilon'$-type contributions are formally of order $h_{li}^4$ in
this highly non-degenerate scenario, other collision terms of order
$h_{li}^4$ may also be significant, such as the scatterings $\Phi L^C
\to \Phi^\dagger L$ and $\Phi \Phi^\dagger \to LL^C$ \cite{MAL,MPl}.
However, the inclusion of these additional effects will not
quantitatively affect our numerical results much, as long as the
constraint in Eq.\ (\ref{hli_bound}) is valid.

In our numerical analysis presented above, we have not taken into
account other effects, which might, to some extend, affect the
resonant condition, given in Eq.\ (\ref{CPcond}).  Apart from the
intrinsic width of a particle resonance, there may be an additional
broadening at high temperatures, due to collisions among particles.
Such effects will contribute terms of order $h^4_{li}$ to the total
$N_i$ widths and are small in general \cite{EWK&SW,Roulet}.  Of most
importance are, however, finite temperature effects on the $T=0$
masses of the particles. Since SM gauge interactions are in kinetic
equilibrium in the heat bath, they can give rise to thermal masses to
the leptons and the Higgs fields \cite{HAW,MEC,CKO}. These thermal
masses are given by
\begin{eqnarray}
\label{thermal}
\frac{m^2_L(T)}{T^2} &=& \frac{1}{32}\, (3g^2\, +\, g'^2)\ \approx\ 0.044
\nonumber\\
\frac{M^2_\Phi (T)}{T^2} &=& 2d\, \Big(\, 1\, -\, \frac{T^2_c}{T^2}\,\Big)\, ,
\end{eqnarray}
where $g$ and $g'$ are the SU(2)$_L$ and U(1)$_Y$ gauge couplings at
the running scale $M_Z$, respectively, and $d = [2M^2_W+M^2_Z+2m^2_t
+M^2_H] / (8v^2)$.  The critical temperature $T_c$ calculated at
one loop may be obtained from \cite{CKO}
\begin{equation}
\label{Tc}
T^2_c\ =\ \frac{1}{4d}\, \Big[\, M^2_H\, -\, \frac{3}{8\pi^2 v^2}\,
(2M^4_W\, +\, M^4_Z\, -\, 4m^4_t )\, -\, \frac{1}{8d\pi^2 v^4}\,
(2M^3_W\, +\, M^3_Z)^2\, \Big]\, . 
\end{equation}
Although the isosinglet heavy neutrinos do not have tree-level
couplings to the SM gauge bosons, the difference between the
thermal mass of $N_i$ and its respective zero-temperature mass will
proceed through Yukawa interactions \cite{HAW}, {\em i.e.}
\begin{equation}
\label{mN(T)}
\frac{m^2_{N_i}(T)\, -\, m^2_{N_i}(0)}{T^2}\ =\ \frac{1}{16}\, |h_{li}|^2\, .
\end{equation}
Such a $T$-dependent shift in the masses of $N_i$ is very small and
may be safely neglected in the mass difference $m^2_{N_1}(T) -
m^2_{N_2}(T)$, which enters the analytic expressions for resonant CP
violation (see, {\em e.g..},\ Eqs.\ (\ref{epsN1}) and (\ref{epsN2})).
Making now use of Eqs.\ (\ref{thermal}) and (\ref{Tc}), the authors in
\cite{CKO} find that $0.5 < M_\Phi (T)/T < 2$, when $M_H$ varies from
60 GeV up to 1 TeV, for $T \gg T_c \approx 200$ GeV. In particular,
one obtains $M_\Phi (T)/T \stackrel{\displaystyle <}{\sim} 0.6$, for
$M_H< 200$ GeV. In this Higgs-mass range, the effective decay
widths of the heavy neutrinos $\Gamma_{N_i}(T)$ will be reduced
relative to $\Gamma_{N_i}(0)$ by 70$\%$ -- 80$\%$, because of
considerable phase-space corrections \cite{MEC}. If $M_H>350$ GeV,
$M_\Phi (T)/T$ is getting bigger than one, which signals the onset of a
non-perturbative regime and pure perturbative methods may not be
sufficient to deal with very heavy Higgs bosons.  For temperatures $T$
near to the critical temperature $T_c$, $M_\Phi (T)/T$ will be smaller
because of the suppression factor $(1-T^2_c/T^2)$ in Eq.\ 
(\ref{thermal}). It appears that low-scale leptogenesis is less
affected by finite temperature effects, even though one can always
choose larger Yukawa couplings to enhance $\Gamma_{N_i}(T)$ in the
light-Higgs scenario. In either case, the resonant phenomenon of
mixing-induced CP violation plays a crucial r\^ole to generate
sufficiently large CP asymmetries.

\setcounter{equation}{0}
\section{Conclusions}

We have studied the impact of the $\varepsilon$- and
$\varepsilon'$-type mechanisms for CP violation on generating the
excess of baryons detected in the Universe. As for the scenario of
baryogenesis, we have considered that out-of-equilibrium $L$-violating
decays of heavy Majorana neutrinos produce an excess in $L$, which is
converted into the observed asymmetry in $B$, through the
$B+L$-violating sphaleron interactions. In Section 2, we have
described minimal extensions of the SM with right-handed neutrinos,
which can predict nearly degenerate heavy Majorana neutrinos without
resorting to a fine tuning of the mass parameters.  Such models may
naturally occur in certain subgroups of SO(10) \cite{Wol/Wyl} or E$_6$
theories \cite{witten}.  In a one-lepton family model, the presence of
two right-handed neutrinos is sufficient to give rise to the
non-trivial CP-violating combination of Yukawa couplings, given in
Eq.\ (\ref{CPinv}). As has been demonstrated in Section 4, our
physical CP asymmetries depend indeed on this CP-odd invariant. In
addition, particular emphasis has been laid on the renormalization of
the Yukawa couplings in the minimally extended model.

In the conventional perturbation theory, the wave-function amplitude
becomes singular, whenever the degenerate limit of the two mixed heavy
neutrinos is considered. Several effective methods have been proposed
to solve this problem, such as diagonalizing the effective Hamiltonian
of the two-heavy-neutrino system \cite{Paschos,LS2}.  The results
obtained with these methods show a resonant enhancement of CP
violation, when the two heavy neutrino masses are getting closer.
Such a resonant CP-violating phenomenon is in line with earlier
studies in \cite{APCP}. Unfortunately, the methods based on
diagonalizing the effective Hamiltonian are not analytic, if the
effective Hamiltonian itself is not diagonalizable \cite{AP}.
Therefore, it is important to compare the results obtained for the CP
asymmetries with a more rigorous field-theoretic approach. In Section
3, we have extended, in an effective manner, the resummation formalism
for particle mixing, which was applied to scatterings in \cite{AP}, to
that of the decays of particles.  The resummed decay amplitudes
possess all the desirable field-theoretic properties and exhibit an
analytic behaviour in the mass degenerate limit.

In Section 4, we have used the afore-mentioned field-theoretic
approach to perform a systematic analysis of the $\varepsilon$ and
$\varepsilon'$ types of CP violation in the $L$-violating decays of
heavy Majorana neutrinos.  For illustration, we have considered
minimal extensions of the SM with two right-handed neutrinos. We have
found that $\varepsilon$-type CP violation is resonantly enhanced up
to order of unity, if the mass splitting of the heavy Majorana
neutrinos is comparable to their widths, as is stated in Eq.\ 
(\ref{CPcond}), and if the parameter $\delta_{CP}$ defined in Eq.\ 
(\ref{dCP}) has a value close to one. In our view, these two necessary
and sufficient conditions for resonant CP violation of order unity
constitute a novel aspect in the leptogenesis scenario, which have not
been pointed out in their most explicit form before.  Taking full
advantage of the mechanism of resonant CP violation, one may consider
scenarios with nearly degenerate heavy Majorana neutrinos at the TeV
scale and all Yukawa couplings being of the same order, which can
still be responsible for the BAU.  In this kinematic range, the
$\varepsilon'$-type contributions are extremely suppressed.  Of
course, the higher the isosinglet Majorana mass is, the less the above
degeneracy is required in order to get sufficiently large CP
violation.  However, even in the weak-mixing limit, {\em i.e.},
$m_{N_1}\ll m_{N_2}$, $\varepsilon$-type CP violation is equally
important with the $\varepsilon'$-type one and therefore should not be
ignored \cite{LS1}. In contrast, only $\varepsilon$-type CP violation
is important in the strong mixing regime, {\em i.e.}, when $m_{N_1} -
m_{N_2} \approx \Gamma_{N_i}$.  Another alternative of having
sufficiently large CP violation for TeV leptogenesis, with
$m_{N_1}={\cal O}(1)$ TeV, is to assume an hierarchic pattern for the
heavy Majorana masses and the Yukawa couplings, {\em e.g.},
$m_{N_2}\gg m_{N_1}$ and $h_{l2}\gg h_{l1}$. Such scenarios were
thoroughly investigated in \cite{MAL,CEV}, and therefore we have not
repeated this analysis here. On the other hand, a wide range of heavy
neutrino masses, $T_c \stackrel{\displaystyle <}{\sim} m_{N_i}
\stackrel{\displaystyle <}{\sim} 10^9$ TeV, is still able to account
for the BAU, even if all Yukawa couplings are of the same order.  We
can hence conclude that the two CP-violating mechanisms under
consideration are, to a great extend, determined from the flavour
structure of the neutrino mass matrix and the flavour hierarchy of the
Yukawa couplings in the model.

In Section 5, we have presented estimates of low-energy constraints on
our minimal model, coming mainly from the electron EDM.  The two-loop
EDM bound derived is found to be not very severe in order to rule out
the leptogenesis scenario and is practically absent if the heavy
Majorana neutrinos are nearly degenerate. In Section 6, we have
briefly discussed the implications of finite temperature effects for
our resonantly enhanced CP violation. Although temperature effects on
the heavy Majorana neutrino masses are very small, because isosinglet
neutrinos interact quite feebly with the Higgs fields in the thermal
bath, the final leptons and Higgs fields acquire appreciable non-zero
thermal masses. As a consequence, there will be a reduction of the
widths of the heavy Majorana neutrinos, relative to their respective
values at $T=0$, due to a phase-space suppression. The size of the
reduction of the $N_i$ widths depends crucially on the
zero-temperature Higgs-boson mass $M_H$. For $M_H< 200$ GeV, the
effective decay width of $N_i$, $\Gamma_{N_i}(T)$, is estimated to be
smaller than $\Gamma_{N_i}(0)$ by $80\%$ at most. This will roughly
lead to an $80\%$ decrease of the CP asymmetries, calculated at zero
temperature.  In this respect, the resonant phenomenon of CP violation
through mixing of heavy neutrinos plays a very important r\^ole for
leptogenesis, since $\varepsilon$-type CP violation can still be large
for heavy-neutrino mass differences comparable to the effective decay
widths $\Gamma_{N_i}(T)$. Finally, further support for the viability
of the resonantly enhanced CP-violating phenomenon is obtained from
solving numerically the Boltzmann equations.  Numerical estimates
reveal that E$_6$-type scenarios, which naturally predict a certain
degree of degeneracy between the heavy Majorana neutrinos, are able to
account for the present excess of baryons. In conclusion, even if all
heavy Majorana neutrinos have masses as low as 1 TeV and all couple to
the lepton and Higgs isodoublets with a universal Yukawa strength,
they can still be responsible for the BAU observed in our epoch, by
means of the resonant mechanism of CP violation presented in this
paper.

\vspace{1.cm}

\noindent {\bf Acknowledgements.} The author wishes to thank Emmanuel
Paschos for useful comments and critical remarks, during his visit at
the University of Dortmund. He also thanks the other members of the
Dortmund group, Jan Weiss and Marion Flanz, for discussions.  Last but
not least, helpful discussions with Zurab Berezhiani, Francisco
Botella, Sacha Davidson, Utpal Sarkar, Mikhail Shaposhnikov and Arkady
Vainshtein are gratefully acknowledged.

\def\theequation{\Alph{section}.\arabic{equation}}
\begin{appendix}
\setcounter{equation}{0}
\section{One-loop analytic expressions}
\indent

In this appendix, we list the analytic expressions for the one-loop
self-energies of the Higgs and fermion fields as well as the one-loop
vertex couplings $\chi^+ lN_i$, $\chi^0\nu_l N_i$ and $H\nu_l N_i$.
Detailed discussion of mixing renormalization for Dirac and Majorana
fermion theories may be found in \cite{KP} and will not be repeated
here. Instead, we present the relations between the wave-function CT's
and unrenormalized self-energies.  Our analytic results will be
expressed in terms of standard loop integrals presented in \cite{HV},
adopting the signature for the Minkowskian metric $g_{\mu\nu} =
\mbox{diag}(1,-1,-1,-1)$ (see also Appendix A of Ref.\ \cite{BAK}).

The Feynman rules used in our calculations may be read off from the
Lagrangian (\ref{LYint}). We first compute the Higgs self-energies
$\chi^-\chi^-$, $\chi^0\chi^0$ and $HH$, shown in Fig.\ 1(d)--(f).
They all found to be equal, {\em viz.}
\begin{eqnarray}
\label{PiHiggs}
\Pi_{\chi^-\chi^-}(p^2)& = & \Pi_{\chi^0\chi^0}(p^2)\ =\ 
\Pi_{HH}(p^2)\nonumber\\
&=& \sum\limits_{l=1}^{n_L} \sum\limits_{i=1}^{n_R}\, 
\frac{|h_{li}|^2}{8\pi^2}
\, \Big[\, m^2_{N_i} B_0(p^2,m^2_{N_i},0)\, +\, p^2\, B_1(p^2,m^2_{N_i},0)\,
\Big]\, .
\end{eqnarray}
Thus, the universality of the divergent parts of the wave functions
$\delta Z_{\chi^-}$, $\delta Z_{\chi^0}$ and $\delta Z_{H}$ is
evident, if one calculates $\delta Z^{div}_\Phi = -\Re
e\Pi'^{div}_\Phi (0)$ from Eq.\ (\ref{PiHiggs}), for all field
components of the Higgs doublet $\Phi$. 

The individual contributions to the one-loop fermionic transitions,
$l'\to l$, $\nu_{l'}\to \nu_l$ and $N_j\to N_i$ are displayed in
Figs.\ 1(g), 1(h) and 1(j), respectively.  Explicit calculation of the
fermion self-energy transitions gives
\begin{eqnarray}
\label{Sigml'l}
\Sigma_{ll'}(\not\! p) &=& - \sum\limits_{i=1}^{n_R}\,
\frac{h_{li}h^*_{l'i}}{16\pi^2} \not\! p\, P_L\, B_1(p^2,m^2_{N_i},0)\, ,\\
\label{Sigmn'n}
\Sigma_{\nu_l\nu_{l'}}(\not\! p) &=& - \sum\limits_{i=1}^{n_R}\,
\frac{1}{32\pi^2}\, \Big[ (h^*_{li}h_{l'i}\not\! p P_R\, +\,
h_{li}h^*_{l'i}\not\! p P_L)\Big(B_1(p^2,m^2_{N_i},0)\, +\,
B_1(p^2,m^2_{N_i},M^2_H)\Big)\nonumber\\
&&+\, m_{N_i}
(h^*_{li}h^*_{l'i} P_R\, +\, h_{li}h_{l'i} P_L)
\Big(B_1(p^2,m^2_{N_i},0)\, -\, B_1(p^2,m^2_{N_i},M^2_H)\Big)\Big],\\
\label{Sigmji}
\Sigma_{N_iN_j}(\not\! p) &=& - \sum\limits_{l=1}^{n_L}\,
\frac{1}{16\pi^2}\, (h^*_{li}h_{lj}\not\! p P_R\, +\,
h_{li}h^*_{lj}\not\! p P_L)\Big(\, \frac{3}{2} B_1(p^2,0,0)\nonumber\\
&& +\, \frac{1}{2} B_1 (p^2,0,M^2_H)\Big) .
\end{eqnarray}
Note that the light-neutrino self-energies
$\Sigma_{\nu_l\nu_{l'}}(\not\! p)$ contain non-vanishing mass terms in
the limit $\not\! p\to 0$.  Even though these contributions vanish
when $M_H\to 0$, they are non-zero at $T=0$, because $M_H\not=0$, and
light neutrinos may hence receive small radiative masses \cite{ZPCAP}.
However, the latter are generally controlled by the mass differences
of the heavy neutrinos and/or the Higgs Yukawa couplings $h_{li}$.  If
the range of parameters relevant for the BAU is considered, as has
been derived in Section 6, possible experimental limits on light
neutrino masses are estimated to be not very restrictive.

The wave-function renormalization constants can now be expressed in
terms of unrenormalized self-energies \cite{KP}. Before doing so, we
first notice that the one-loop $f_j\to f_i$ transitions calculated above
between the fermions $f_i=l,\ \nu_l,\ N_i$ have the generic form
\begin{equation}
\label{Sigma}
\Sigma_{ij}(\not\! p)\ =\ \Sigma^L_{ij} (p^2)\not\! p P_L\, +\,
\Sigma^R_{ij} (p^2)\not\! p P_R\, +\, \Sigma^M_{ij}(p^2) P_L\, +\,
\Sigma^{M*}_{ji}(p^2) P_R\, ,
\end{equation}
where only dispersive parts are considered.  If the transitions are
between Majorana fermions in Eq.\ (\ref{Sigma}), one then has the
extra properties $\Sigma^L_{ij} (p^2)= \Sigma^{R*}_{ij} (p^2)$ and
$\Sigma^M_{ij}(p^2)=\Sigma^M_{ji}(p^2)$. Adapting the results of
\cite{KP} to our model, we obtain the wave-function CT's 
\begin{eqnarray}
\label{dZfii}
\delta Z^f_{ii} &=& -\Sigma^L_{ii}(m^2_i)\, -\, 2m^2_i\Sigma^L_{ii}{}'(m^2_i)
\, -\, 
m_i\Big[ \Sigma^M_{ii}{}'(m^2_i)+\Sigma^{M*}_{ii}{}'(m^2_i)\Big]\nonumber\\
&&+\, \frac{1}{2m_i}\Big[ \Sigma^M_{ii}(m^2_i)-\Sigma^{M*}_{ii}(m^2_i)\Big],
\end{eqnarray}
and, for $i\not= j$,
\begin{equation}
\label{dZfij}
\delta Z^f_{ij}\ =\ \frac{2}{m^2_i-m^2_j}\Big[
m^2_j\Sigma^L_{ij}(m^2_j)\, +\, m_im_j\Sigma^{L*}_{ij}(m^2_j)\, +\,
m_i\Sigma^M_{ij}(m^2_j)\, +\, m_j\Sigma_{ij}^{M*}(m^2_j)\Big].
\end{equation}
The wave-function renormalization of charged leptons may be recovered
from Eqs.\ (\ref{dZfii}) and (\ref{dZfij}), if one drops all terms
depending on $\Sigma^M_{ij}(p^2)$ and its derivative. At this point,
it is important to remark that there will be additional contributions
to our wave-function CT's of the Higgs and fermion fields from the
gauge sector of the SM \cite{BAK}. For example, there are non-zero
thermal mass effects on the particles involved in the loop, which may
further break the wave-function universality of the different
components of the Higgs doublet \cite{HAW,MEC}. As has been discussed
in Section 2, this should not pose any problem to the Yukawa coupling
renormalization, as long as all differences of the kind $\delta
Z_{\chi^-} -\delta Z_H$ or $\delta Z_{\chi^-} -\delta Z_{\chi^0}$ are
UV finite.

We will now present analytic results for the one-loop corrections
to the vertices $\chi^\pm l^\mp N_i$, $\chi^0 \nu_l N_i$ and $H \nu_lN_i$,
keeping all $M_H$-dependent mass terms. From Figs.\ 1(a)--(c), we
observe that the one-loop coupling  $\chi^\pm l^\mp N_i$ proceeds 
via a $\Delta L=2$ mass insertion only, whereas the couplings
$\chi^0 \nu_l N_i$ and $H \nu_lN_i$ can occur through both $L$-conserving
and $L$-violating interactions. More explicitly, we have
\begin{eqnarray}
\label{chi_N}
i{\cal V}_{\chi^+l^-N_i} &=& -i \bar{u}_l P_R u_{N_i}\, 
\sum\limits_{l'=1}^{n_L}\sum\limits_{j=1}^{n_R}\,
\frac{m_{N_i}m_{N_j}}{16\pi^2}\ h^*_{l'i}h_{l'j}h_{lj} 
\nonumber\\
&&\times\, \Big[ C_0(0,0,m^2_{N_i},0,m^2_{N_j},0)\, +\, 
C_{12}(0,0,m^2_{N_i},0,m^2_{N_j},0)\, \Big] ,\\
\label{chi0N}
{\cal V}_{\chi^0\nu_l N_i} &=& -\, \bar{u}_l P_R u_{N_i}\, 
\sum\limits_{l'=1}^{n_L}\sum\limits_{j=1}^{n_R}\,
\Big\{\, \frac{m_{N_i}m_{N_j}}{32\sqrt{2}\pi^2}\
h^*_{l'i}h_{l'j}h_{lj} \nonumber\\
&&\times\, \Big[C_0(0,0,m^2_{N_i},0,m^2_{N_j},0)\, +\, 
C_0(0,0,m^2_{N_i},M^2_H,m^2_{N_j},0)\nonumber\\
&&+C_{12}(0,0,m^2_{N_i},0,m^2_{N_j},0)\, +\, 
C_{12}(0,0,m^2_{N_i},M^2_H,m^2_{N_j},0)\Big]\nonumber\\
&&+\frac{1}{32\sqrt{2}\pi^2}\ h_{l'i}h^*_{l'j}h_{lj}
\nonumber\\
&&\times \Big[\, m_{N_i}m_{N_j}\Big(C_{12}(0,0,m^2_{N_i},0,m^2_{N_j},0)\, -\, 
C_{12}(0,0,m^2_{N_i},M^2_H,m^2_{N_j},0)\Big)\nonumber\\
&&-\, M^2_H C_0 (0,0,m^2_{N_i},M^2_H,m^2_{N_j},0) \Big]\, \Big\}, \\
\label{HN}
-i{\cal V}_{H\nu_lN_i} &=& i\bar{u}_l P_R u_{N_i}\,  
\sum\limits_{l'=1}^{n_L}\sum\limits_{j=1}^{n_R}\,
\Big\{\, \frac{m_{N_i}m_{N_j}}{32\sqrt{2}\pi^2}\ h^*_{l'i}h_{l'j}h_{lj}
\nonumber\\
&&\times\, \Big[C_0(0,0,m^2_{N_i},0,m^2_{N_j},0)\, +\, 
C_0(0,0,m^2_{N_i},M^2_H,m^2_{N_j},0)\nonumber\\
&&+C_{12}(0,0,m^2_{N_i},0,m^2_{N_j},0)\, +\, 
C_{12}(0,0,m^2_{N_i},M^2_H,m^2_{N_j},0)\Big]\nonumber\\
&&+\frac{1}{32\sqrt{2}\pi^2}\ h_{l'i}h^*_{l'j}h_{lj}
\nonumber\\
&&\times \Big[\, m_{N_i}m_{N_j}\Big(C_{12}(0,0,m^2_{N_i},0,m^2_{N_j},0)\, -\, 
C_{12}(0,0,m^2_{N_i},M^2_H,m^2_{N_j},0)\Big)\nonumber\\
&&-\, M^2_H C_0 (0,0,m^2_{N_i},M^2_H,m^2_{N_j},0) \Big]\, \Big\} .
\end{eqnarray}
As can be readily seen from Eqs.\ (\ref{chi0N}) and (\ref{HN}), the
part of the couplings $\chi^0 \nu_l N_i$ and $H \nu_lN_i$ that
conserves lepton number is UV finite and vanishes identically in the
massless limit of the Higgs boson.

\end{appendix}

\end{document}